\documentclass[prb,twocolumn,showpacs,preprintnumbers,amsmath,amssymb,floatfix]{revtex4}
\usepackage{graphicx}
\usepackage{bm}
\usepackage{float}

\newcommand \be{\begin{eqnarray}}
\newcommand \ee{\end{eqnarray}}
\newcommand \ba{\begin{align}}
\newcommand \eea{\end{align}}
\newcommand {\ket}[1]{|#1\rangle}
\newcommand {\bra}[1]{\langle #1|}

\newcommand {\p}[1]{\partial_{#1}}

\newcommand \V{\vec}
\newcommand \uu{\uparrow}
\newcommand \dd{\downarrow}

\begin{document}
           \csname @twocolumnfalse\endcsname
\title{Kinetic theory of spin-polarized systems in electric and magnetic fields with spin-orbit coupling: I. Kinetic equation and anomalous Hall and spin-Hall effects}
\author{K. Morawetz$^{1,2,3}$ 
}
\affiliation{$^1$M\"unster University of Applied Sciences,
Stegerwaldstrasse 39, 48565 Steinfurt, Germany}
\affiliation{$^2$International Institute of Physics (IIP)
Federal University of Rio Grande do Norte
Av. Odilon Gomes de Lima 1722, 59078-400 Natal, Brazil
}
\affiliation{$^{3}$ Max-Planck-Institute for the Physics of Complex Systems, 01187 Dresden, Germany
}

\begin{abstract}
The coupled kinetic equations for density and spin Wigner functions are derived including spin-orbit coupling, electric and magnetic fields, selfconsistent Hartree meanfields suited for SU(2) transport. The interactions are assumed to be with scalar and magnetic impurities as well as scalar and spin-flip potentials among the particles. The spin-orbit interaction is used in a form suitable for solid state physics with Rashba or Dresselhaus coupling, graphene, extrinsic spin-orbit coupling, and effective nuclear matter coupling. The deficiencies of the two-fluid model are worked out consisting of the appearance of an effective in-medium spin precession. The stationary solution of all these systems shows a band splitting controlled by an effective medium-dependent Zeeman field. The selfconsistent precession direction is discussed and a cancellation of linear spin-orbit coupling at zero temperature is reported. The precession of spin around this effective direction caused by spin-orbit coupling leads to anomalous charge and spin currents in an electric field. Anomalous Hall conductivity is shown to consists of the known results obtained from the Kubo formula or Berry phases and a new symmetric part interpreted as an inverse Hall effect. Analogously the spin-Hall and inverse spin-Hall effects of spin currents are discussed which are present even without magnetic fields showing a spin accumulation triggered by currents. The analytical dynamical expressions for zero temperature are derived and discussed in dependence on the magnetic field and effective magnetizations. The anomalous Hall and spin-Hall effect changes sign at higher than a critical frequency dependent on the relaxation time.  
\end{abstract}
\maketitle
%

\section{Introduction}

\subsection{Motivation and outline}

The interest in spin transport has regained a renaissance due to the promising application as next generation information storage. The experiments have reached such high precession that single spin transport and spin wave processes can be resolved and investigated in view of applications to new nanodevices where spin-field transistors are proposed \cite{PhysRevLett.90.146801}. 
Many interesting effects have been reported such as anomalous spin segregation in weakly interacting $^6$Li in a trap \cite{Natu09} which effect has been described by the meanfield, spin-Hall effects and spin-Hall nano-oscillators \cite{DUG14}. Different spin transport effects have led to the spin current concept \cite{Sonin:2010} which tries to summarize these effects with respect to the current.

The spin-orbit coupling has lead to the idea of the spin-Hall effect \cite{W03,Sch06,Dyakonov:2009} which was proposed \cite{DP71,DP71a} and first observed in bulk $n$-type semiconductors \cite{Ka04} and in 2D heavy-hole systems \cite{Wu05}. 
Spin and particle currents are coupled such that one observes an accumulation of transverse spin current near the edges of the sample. 
The spin conductance has been measured in mesoscopic cavities \cite{Adagideli:2009} to extract the part due to spin-orbit coupling. Within the anomalous Hall effect first described in Ref. \cite{Kar54} the spin polarization takes over the role of a magnetic field and creates a current contribution \cite{Sinitsyn:2007,Kovalev:2009,Liu05}. This effect occurs when time-symmetry is broken \cite{Nagaosa:2010} and is related to the inverse spin Hall effect \cite{Schwab:2010}. For the latter one spin-orbit coupling takes the role of an additional electric field.

Anisotropic magnetoresistance together with the anomalous Hall coefficients have been measured and attributed
to spin-orbit coupling \cite{Bibes:2005,Gould:2007} and treated in quantum wires \cite{Hattori:2010} and mesoscopic rings \cite{Souma:2005}. The lattice structure causes strong dependencies on the transport direction \cite{Liu:2007}.

One distinguishes between extrinsic and intrinsic spin-Hall effects. The extrinsic is due to spin-dependent scattering by mixing of spin and momentum eigenstates. The intrinsic effect is an effect of the momentum-dependent internal magnetic field due to spin-orbit coupled band structures. This leads to a spin splitting of the energy bands in semiconductors due to lacking of inversion symmetry. Most observations are performed with the extrinsic \cite{Ka04,Sih:2005,VT06,SHMTI08} and only some for the intrinsic spin-Hall effect \cite{Wu05,Br10}.  There are different model treatments of intrinsic \cite{Fujita:2009,Hankiewicz:2004} and extrinsic spin-Hall effects \cite{Gradhand:2010} sometimes using Berry curvatures \cite{Nikolic:2006,Gosselin:2009,Nagaosa:2010,PhysRevB.84.075113}, the Landauer formula \cite{Nikolic:2006} or even relativistic treatments \cite{Tse:2005,Lowitzer:2010}. A theoretical comparison of the relativistic approach with the Kubo formula is found in \cite{CB01}.
A detailed discussion of possible occurring spin-orbit couplings in semiconductor bulk structures and nanostructures can be found in \cite{WJW10} and in the book \cite{W03}. 
For the Rashba coupling and quadratic dispersion in disordered two-dimensional systems it has been shown that the spin-Hall effect vanishes \cite{Liu06,Kh06,Mal'shukov:2005}. This is not the case if magnetic scatters are considered \cite{Inoue:2006}. The intrinsic anomalous Hall effect is treated also in Ref. \cite{Streda:2010} and in disordered band ferromagnets \cite{Wolfle:2006}. 

The (pseudo)spin-Hall effect in graphene is currently a very heavily investigated field \cite{Kane:2005} reporting also the anomalous Hall effect in single-layer and bilayer graphene \cite{PhysRevB.82.161414,PhysRevB.83.155447} and which is treated like a spin-orbit coupled system \cite{Tk09}. Recently even spin-orbit coupled Bose-Einstein condensates have been realized \cite{LGS11}.

The main motivation of the present paper is to derive in an unambiguous way 
a kinetic equation of interacting spin-polarized fermions
including magnetic and electric fields with spin-orbit coupling. Normally one finds all four problems treated separately in the literature. 
First, there exists a vast literature to derive kinetic equations with spin-polarized electrons \cite{LhLa82a,LhLa82b,Me89,Me91,JeMu88,JeMu89,NaTa89,NaTa91,MeMu92,Bas91,HaMo92,Bas00}. 
Second, other quantum kinetic approaches focus exclusively on the spin-orbit coupling \cite{E90,Raimondi:2006,RS10,SB05}. 
Third, the transport in high and low magnetic fields itself is involved due to precession motions of charged particles and treated approximately \cite{Be58,KD98,Shiz07,KB09,ZMB10,LB10,WZT05}. 
Fourth, the interaction with scalar and magnetic impurities requires a certain spin-coupling which is important for transport effects in ferromagnetic materials \cite{SCC89,O98,CH98}.

Here we will combine all four difficult problems into a unifying quantum kinetic theory. First we restrict ourselves to approximate the many-body interactions by the mean field and a conserving relaxation time. The outlined formalism is straight forward to derive proper collision integrals as done in the literature \cite{JeMu88,Me89,MeSt95,NaTa91,Kailasvuori:2009,Kailasvuori:2010}. The reasons to consider once again the lowest-order many-body approximation is twofold. On one side during the derivation of the proper kinetic equations it turned out that even on the meanfield level all four effects together create additional terms when considered on a common footing not known so far. On the other side, with a meanfield quantum kinetic equation including all these effects we have the possibility to linearize with respect to an external perturbation and to obtain  in this way the response function in the random phase approximation (RPA). As a general rule, when a lower-level kinetic equation is linearized, a response of higher-order many-body correlations is obtained\cite{Mc02}.  

Most treatments of the response function use approximations already at the beginning and concentrate only on specific effects, such asthe diffusive regime \cite{Bleibaum:2006,Nikolic:2006} or currents \cite{Mishchenko:2004}. We will explicitly work out these response functions in the second paper of this series. Here in the first paper we want to focus on the derivation of the quantum kinetic equation including all these effects as transparently as possible. With the aim to derive the RPA response we concentrate on the correct mean-field formulation and restrict ourselves to a relaxation-time approximation of the collision integral as a first step. This relaxation time will be understood with respect to a local equilibrium which accounts for local conservation laws \cite{Mer70,D75,HPR93,BC10,Ms12}. Though it can be derived from Boltzmann collision integrals, this relaxation time approximation omits quantum interference effects such as weak localization due to disorder, for an adequate treatment of these effects see, e.g., \cite{ZNA01}. 

The outline of this first paper is as follows. After explaining the basic notation we present in the second section the phenomenological two-fluid model and show the insufficiency due to the missing self-consistent precession direction. We present an educative guess for the proper kinetic equation from the demand of SU(2) symmetry. Interestingly, this leads already to the correct form of the kinetic equation except for four parameters which have to be derived microscopically in Sec. III. We use the nonequilibrium Green's function technique in the notation of Langreth and Wilkins \cite{LW72}. The obtained kinetic equations lead to a unique static solution which shows the splitting of the band due to spin-orbit interference. The anomalous current is shown to cancel the normal one in the stationary state pointing to the importance of the anomalous currents when the balance is disturbed like in transport. The selfconsistent precession direction is calculated explicitly for zero temperature and linear spin-orbit coupling. In Sec. IV we compare the derived kinetic equation with the guessed one of Sec. II determining the remaining open parameters. The anomalous Hall and spin-Hall effects are calculated in Sec. IV and appear in agreement with other approaches using the Kubo formula or helicity basis. We obtain a dynamical symmetric contribution interpreted as inverse spin-Hall and inverse Hall effects. Analytical expressions are discussed for the dynamical conductivities at zero temperature and linear spin-orbit coupling. A summary concludes the first paper of this series.

\subsection{Basic notation}

The spin of a fermion $\frac \hbar 2 \V \sigma$ with the Pauli matrices $\V \sigma$ as an internal degree of freedom analogously to circular motion leads to the elementary magnetic moment for the spin in terms of the Bohr magneton $\mu_B$,
\be
\hat {\V m}=g{e \over 2 m_e}\frac \hbar 2 \V \sigma=\frac g 2 \mu_B \V \sigma 
\ee
with the anomalous gyromagnetic ration $g\approx 2$ for electrons. 
If there are many fermions with densities $n_\pm$ of spins parallel/antiparallel to the magnetic field, the total magnetization is $m_z=g \mu_B s_z$ with the polarization $s_z=(n_+-n_-)/2$
We want to access the density and polarization density distributions and use therefore four Wigner functions
\be
\hat \rho(\V x,\V p)=f(\V x, \V p)+\V \sigma\cdot \V g(\V x, \V p)=
\begin{pmatrix}
f+g_z &g_x-i g_y\cr g_x+i g_y&f-g_z
\end{pmatrix}
\label{rhofg}
\ee
with the help of which the density and polarization density are given by
\be 
\sum_p f=n(\V x),\qquad
\sum_p \V g=\V s(\V x)
\label{quasin}
\ee
where $\sum\limits_p=\int d^Dp/(2\pi\hbar)^D$ for $D$ dimensions and the magnetization density becomes
$\V M(\V x)=g  \mu_B \V s(\V x)$. The advantage is that we can describe any direction of magnetization density created by microscopic correlations which will be crucial in this paper.
Sometimes one finds the probability distribution of spin up/down in the direction $\V e$ of the mean polarization by the spin projection or a twofold additional spin variable \cite{ZMB10,LB10} is used. Since there are inversion formulas \cite{M88} all these approaches should be equivalent.
However, we prefer the presentation of the Wigner function in terms of the scalar and vector parts (\ref{rhofg}) since the coupling between these functions bear clear physical meaning which is somewhat buried in the sometimes used super distribution. Moreover the Wigner functions (\ref{quasin}) yield directly the total density and the spin-polarizations.

\subsection{Spin-orbit coupling}

Any spin-orbit coupling used in different fields, say plasma systems, semiconductors, graphene or nuclear physics can be recast into the general form 
\be
H^{\rm s.o.}=A(\V p) \sigma_x -B(\V p) \sigma_y +C(\V p) \sigma_z=\V b \cdot \V \sigma
\label{so}
\ee 
with a momentum-dependent $\V b$ illustrated in table~\ref{ab} and which can become space and also time-dependent in nonequilibrium. Also the Zeeman term is of this form. 
The time reversal invariance of spin current due to spin-orbit coupling 
requires that the coefficients $A(p)$ and $B(p)$ be odd functions of the momentum $k$ and therefore such couplings have no spatial inversion symmetry.
Also 3D systems of spin-1/2 particles can be recast into the form (\ref{so}).
Let us shortly discuss different realizations since we want to treat all of them within the quantum kinetic theory.

\subsubsection{Extrinsic spin-orbit coupling}

First we might think on the direct spin-orbit coupling as it appears due to expansion of the Dirac equation where only the Thomas term is relevant
\be
\V \sigma \cdot {i e \hbar^2\over 8 m^2 c^2} \left (\p R\!\times\! \V E\!-\!{2 i\over \hbar} \V E\!\times\! \V p\right )
&\approx &
\lambda^2 \V \sigma \cdot ({\V p \over \hbar}\times \nabla V)
\label{dirac}
\ee
with $\lambda^2=\hbar^2/4 m_e^2 c^2\approx 3.7\times 10^{-6}$\AA$^2$ for 
electrons. The electric field is not an external
one, but e.g. created by the nucleus $\V E=-\V \nabla V$, and is called extrinsic spin-orbit coupling. The spin-orbit coupling mixes different momentum states and is coupled to inhomogeneities in the material. The matrix elements of the spin-orbit potential reads
\be
\langle \V p_2 |V^{\rm s.o.}|\V p_1 \rangle&=&{i\lambda^2\over \hbar^2} V(\V p)(\V p \times \V \sigma )\cdot \V q
\label{100}
\ee 
with the center-of-mass momentum $\V q=(\V p_1+\V p_2)/2$ and $\V p=\V p_1-\V p_2$.
Any such spin-orbit coupling possesses the general structure (\ref{so}).

\subsubsection{Intrinsic spin-orbit coupling in semiconductors}

For direct gap cubic semiconductors such as GaAs the form (\ref{so}) of spin-orbit coupling arises by coupling of the s-type conductance band to p-type valence bands. With in the 8$\times$8 Kane model the third-order perturbation theory\cite{W03} yields $\lambda=P^2/3[1/E_0^2-1/8(E_0+\Delta_0)^2]$  with the gap $E_0$ and the spin-orbit splitting $\Delta_0$ between the $J=3/2$ and $J=1/2$ hole bands and a matrix element $P$. For GaAs one finds $\lambda=5.3$\AA$^2$ which shows that in n-type GaAs the spin-orbit coupling is six orders of magnitude stronger than in vacuum and has an opposite sign \cite{EHR05}. The cubic Dresselhaus spin-orbit corrections are usually neglected since they are small and does not appear in the 8$\times$8 model. Therefore the spin-orbit coupling is considered to come from the potential of the driving field and the impurity centers.

In a GaAs/AlGaAs quantum well there can be two types of spin-orbit couplings that are linear in momentum. One considers a narrow quantum well in the $\V n=[001]$ direction.
The linear Dresselhaus spin-orbit coupling is due to the bulk inversion asymmetry of the zinc-blende type lattice. It is proportional to the kinetic energy of the electron's out-of-plane motion and  decreases therefore quadratically with increasing well width. In lowest-order momentum one obtains
\be
H^{\rm s.o.}_\mathrm{D} &=& {\beta_D\over \hbar}  (-p_y \sigma^y+p_x \sigma^x)
\ee
again of the form (\ref{so}) with $\V b=\beta_D (-p_x,p_y,0)/\hbar$.

The Rashba spin-orbit coupling (SOC) 
\be
H^{\rm s.o.}_\mathrm{R} &=& {\beta_R\over \hbar}  (-p_x \sigma^y+p_y \sigma^x)={\beta_R\over \hbar}  \V \sigma \cdot (\V p\times \V n)
\label{Rashba}
\ee
 is finally due to structure inversion asymmetry and the strength can be tuned by a perpendicular electric field, for
 example by changing the doping imbalance on both sides of the quantum
 well. The Rashba coupling is again of the form (\ref{so}) with $\V b=\beta_R (p_y,-p_x,0)/\hbar$.
Note that the Rashba  SOC has winding number +1 as the momentum direction winds around once in momentum space, whereas the linear Dresselhaus has the opposite winding -1 \cite{MLRK12}.

There are further types of spin-orbit expansion schemes for quasi-2D systems such as cubic Rashba and cubic Dresselhaus expansions whose competing interplay is treated too\cite{Chang:2009}. These terms including wurtzite structures \cite{Chen09} all together can be recast into the form of (\ref{so}) and seen in table \ref{ab}. 

\begin{table}
\caption{\label{ab} Selected 2D and 3D systems with the Hamiltonian described by (\ref{so}) taken from \cite{Chen09,cserti06} }
\begin{align*}
&
\begin{array}{llll}
{\rm 2D-system} & A(p) & B(p) & C(p)\cr
\hline
{\rm Rashba} & \beta_R p_y & \beta_R p_x &\cr
{\rm Dresselhaus} [001] & \beta_D p_x & \beta_D p_y &\cr
{\rm Dresselhaus} [110] & \beta p_x & -\beta p_x &\cr
{\rm Rashba-Dresselhaus} & \beta_R p_y-\beta_D p_x & \beta_R p_x-\beta_D p_y &\cr
{\rm cubic Rashba (hole)} & i{\beta_R \over 2}(p_-^3-p_+^3)&{\beta_R \over 2}(p_-^3+p_+^3) &\cr
{\rm cubic Dresselhaus} & \beta_D p_x p_y^2& \beta_D p_y p_x^2 &\cr
{\rm Wurtzite type} & (\alpha +\beta p^2 ) p_y & (\alpha+\beta p^2 ) p_x&\cr
{\rm single-layer graphene} & v p_x & -v p_y&\cr
{\rm bilayer graphene} & {p_-^2+p_+^2\over 4m_e} & {p_-^2-p_+^2\over 4m_e i}& 
\end{array}
\end{align*}
\begin{align*}
&
\begin{array}{llll}
{\rm 3D-system} & A(p) & B(p)& C(p) \cr
\hline
{\rm bulk}\, {\rm Dresselhaus} & p_x(p_y^2-p_z^2) & p_y(p_x^2-p_z^2) &p_z(p_x^2-p_y^2) \cr
{\rm Cooper pairs} & \Delta & 0 & {p^2\over 2 m} -\epsilon_F \cr
{\rm extrinsic} &&&\cr
 \beta={i\over \hbar}\lambda^2 V(p) &q_y p_z-q_z p_y&q_z
p_x-q_x p_z &q_x p_y-q_y p_x \cr
{\rm neutrons \, in\,  nuclei} &&&\cr
 \beta={i} W_0 (n_n+{n_p\over 2}) &q_z p_y-q_y p_z&q_x p_z -q_z
p_x&q_y p_x -q_x p_y
\end{array}
\end{align*}
\end{table}

\subsubsection{Spin-orbit coupling in graphene}

Solving the  tight-binding model on the honeycomb lattice including only nearest neighbor hopping gives an effective two-band Hamiltonian for the Bloch wave function which can be considered as coupled pseudo-spins.
The interband coupling of these different bands in graphene leads to the spin-orbit coupling of the form (\ref{so}). 

\subsubsection{Spin-orbit coupling in nuclear matter}

In nuclear matter the spin-orbit interaction is strong in heavy elements and the reason for magic numbers. Shell structures cannot be described properly without consideration of the spin-orbit interaction coming from the tensor part of the nuclear forces \cite{BBDHLM09}. The effective spin-orbit coupling in nuclear matter with neutrons and protons can be considered as a meanfield expression due to schematic Skyrme forces and is expressed in the Rashba form (\ref{Rashba}) which reads for neutrons \cite{YJLW14}
\be
H^{\rm s.o.}=-{W_0\over 2}\V \sigma [\V p\times (\V \nabla n_p+2\V \nabla n_n)]
\ee
and interchanging $n_p\leftrightarrow n_n$ for protons. One sees that in this Hartree-Fock expression the effective direction $\V n$ of Rashba form (\ref{Rashba}) is given by the gradient of the densities. Therefore the structure appears as in extrinsic spin-orbit coupling. The coupling constant is a matter of debate and dependent on the used density functional \cite{ZDSW08}. Further terms can be considered if more involved Skyrme potentials are used leading to additional current coupling\cite{PF94,LBBDM07}. 
It is expected that the spin-orbit coupling plays an important role in heavy-ion reactions \cite{XB13} and in rare isotopes.

\section{Phenomenological kinetic equation}

\subsection{Deficiencies of two fluid model}

Now we consider the widely used two-fluid model
\cite{FC68,Bala14} developed from the two-current conduction in iron
\cite{CFP67}. It consists of a distribution for spin-up $f_\uu$ and spin-down
$f_\dd$ parts. Until recently it was used to explain even anisotropic
magnetic resistance \cite{RHA14} with its limits observed there. 
Indeed we will show that an important part is missing in this model. Besides
the two distribution functions we have to consider the direction of the mean
spin or polarization. As we will see this leads to a third equation which is
silently overlooked in these models. From general SU(2) symmetry
considerations one can already conclude that this part is missing in the
two-fluid model and how its form should appear. The detailed derivation 
will be performed in the next section. Here we repeat briefly the two-fluid model
and develop the missing parts from general physical grounds.

We start with the linearized coupled kinetic equations for the two components
\be
&&\partial_t \delta f_\uu+{\V p\over m_e}\V \partial_r \delta f_\uu-\V \partial_r (U^{\rm
  ext}+U_{\uu\uu} \delta n_\uu+U_{\uu\dd} \delta n_\dd)\V \partial_p
f_\uu^0\nonumber\\
&&=-{\delta f_\uu\over \tau_{\uu\uu}}-{\delta f_\uu-\delta f_\dd\over
  \tau_{\uu\dd}}\nonumber\\
&&\partial_t \delta f_\dd+{\V p\over m_e}\V \partial_r \delta f_\dd-\V \partial_r (U^{\rm
  ext}+U_{\dd\dd} \delta n_\dd+U_{\dd\uu} \delta n_\uu)\V \partial_p
f_\dd^0\nonumber\\
&&=-{\delta f_\dd\over \tau_{\dd\dd}}-{\delta f_\dd-\delta f_\uu\over
  \tau_{\dd\uu}}
\label{twofluid}
\ee
with the external electric field $e\V E=-\V \partial_r U^{\rm ext}$. The relaxation due to collisions is considered as relaxation times with
respect to the same kind of particle $\tau_{\uu\uu}$ or $\tau_{\dd\dd}$ and with respect to the other sort which is
described by the cross relaxation time $\tau_{\uu\dd}=\tau_{\dd\uu}$ due to
symmetric collisions. For later use we have added also the
spin-dependent meanfields ${\cal U_i}$ and their linearization $U_{ij}=\partial {\cal U}_i/\partial n_j$ with respect to the densities
$
n_{\uu,\dd}=\sum_p f_{\uu,\dd}
$.

\begin{figure}
\includegraphics[width=8cm]{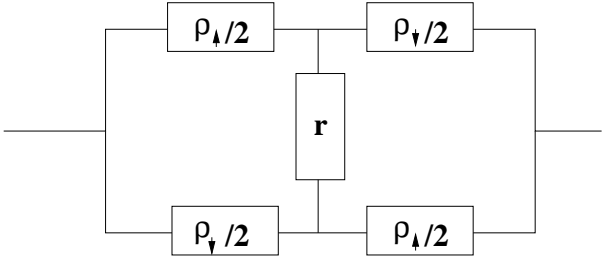}
\caption{Resistor scheme of (\ref{rr}) illustrating the spin mixing.\label{resist}}
\end{figure}

Multiplying (\ref{twofluid}) with $\V p$ and integrating together with Fourier transform
$\partial_t \to -i\omega$ and $\V \partial_r\to i \V q$ leads to the coupled equations
for the currents in lowest-order wavevector $\V q$
\be
(\rho_\uu+r_{\uu\dd}) \delta J_\uu-r_{\uu\dd}\delta J_\dd&=&E
\nonumber\\
-r_{\dd\uu}\delta J_\uu+(\rho_\dd+r_{\dd\uu}) \delta J_\dd&=&E
\label{jj}
\ee
where we have introduced the partial and crossed resistivities [$(i,j)=\uu,\dd$]
\be
{1\over \rho_i}&=&\sigma_i={n_i e^2\tau_{ii}\over m_e
  (1-i\omega\tau_{ii})},\qquad
{r_{ij}}={m_e\over n_i e^2\tau_{ij}}.
\ee
The partial currents (\ref{jj}) are easily solved and one obtains the total
resistivity
\be
\rho={E\over \delta J_\uu+\delta J_\dd}={\rho_\uu\rho_\dd+\rho_\uu
  r_{\uu\dd}+\rho_\dd r_{\uu\dd}\over \rho_\uu+\rho_\dd+2
  (r_{\uu\dd}+r_{\dd\uu})}.
\label{rr}
\ee
Assuming further that $r=r_{\dd\uu}=r_{\uu\dd}$ this resistivity allows an
interpretation as composed resistivity illustrated in figure \ref{resist}
which shows the role of the cross scattering between different species known as
spin mixing.

Despite the successful application of this model \cite{C03} it has an important
inconsistency which is not easy to recognize. We therefore rewrite the kinetic
equations (\ref{twofluid}) into a form for the total density distribution and
total density
\be
f=\frac 1 2 (f_\uu+f_\dd);\quad n=\frac 1 2 (n_\uu+n_\dd)
\ee
and the polarization distribution and total spin
\be
g=\frac 1 2 (f_\uu-f_\dd);\quad s=\frac 1 2 (n_\uu-n_\dd)
\ee
which reads (Fourier transform $\V \partial_r \to i\V q$)
\ba
&\partial_t \delta f+{i \V q \V p\over m_e} \delta f+e\V E \V \partial_p f^0-M_f=-{\delta f\over \tau_+}-{\delta g\over
  \tau_-}
\nonumber\\
&\partial_t \delta g+{i \V q \V p\over m_e}\delta g+e\V E\V \partial_p g^0-M_g=-{\delta f\over \tau_-}-{\delta g\over
  \tau_+}-2{\delta g\over \tau_D}
\label{fg}
\end{align}
with the meanfield parts abbreviated as
\ba
M_f&= i \V q \V \partial_p f^0 (V_1 \delta n+V_3\delta s)
+i \V q \V \partial_p g^0(V_2 \delta n+V_4\delta s)
\nonumber\\
M_g&=i \V q\V \partial_p g^0(V_1 \delta n+V_3 \delta s)
+i \V q\V \partial_p f^0(V_2 \delta n+V_4 \delta s).
\end{align}
Here it was convenient to introduce the relaxation times
\be
{1\over \tau_\pm}=\frac 1 2 \left ({1\over \tau_{\uu\uu}}\pm{1\over
    \tau_{\dd\dd}}\right );\quad \tau_D=\tau_{\uu\dd}=\tau_{\dd\uu}
\label{relaxnew}
\ee
and the meanfield potentials
\be
V_{1/2}&=&U_{\uu\uu}+U_{\uu\dd}\pm (U_{\dd\uu}+U_{\dd\dd})\nonumber\\
V_{3/4}&=&U_{\uu\uu}-U_{\uu\dd}\pm (U_{\dd\uu}-U_{\dd\dd}).
\label{potmean}
\ee

The crucial point is now that the polarization or total spin has a direction
$\V e$ which means we have to consider the vector quantity
$
\V g=g\V e
$ which translates into two parts when linearized $\delta \V g=g\delta \V e+\V e
\delta g$. Only the second part is obviously covered by the second equation of
(\ref{fg}). The equation for $\delta \V e$ remains undetermined so
far. 

\subsection{Educated guess from SU(2) symmetry}

However from general consideration of SU(2) symmetry we can infer the
form in which this missing equation will appear and which we will derive in the next section from microscopic theory.

It is convenient to write both scalar distribution $f$ and vector distribution
$\V g$ together in spinor form (\ref{rhofg}). Then any collision integral and
therefore any kinetic equation must be possible to write as
commutator and anticommutator in spin-space where the forms
\be
\frac 1 2 \left [\delta \hat \rho,a+\V \sigma\cdot \V b\right
]_+&=&{a \delta f}
+\V b\cdot \delta \V g+\V \sigma\cdot \left (a {\delta \V g}+\V b
\delta f\right )\nonumber\\
\frac 1 2 \left [\delta \hat \rho,a+\V \sigma\cdot \V c\right ]_-&=&
i\V \sigma \cdot (\delta \V g\times \V c)
\label{rules1}
\ee
can appear making use of $(\V a\cdot \V \sigma)(\V b\cdot \V \sigma)=\V a\cdot \V b+i\V \sigma\cdot (\V a\times \V b)$.
Therefore the expected kinetic equation reads
\ba
&\partial_t \delta \hat \rho\!+\!{i \V p \V q \over m_e}\delta \hat \rho\!+\!e\V E\V \partial_p \hat
\rho^0
\nonumber\\
&-\frac i 2 \left [\V q \V {\p p} \hat \rho^0,a+\V b  \V \sigma
\right ]_+
-\frac i 2 \left [\V q\V \partial_p \hat \rho^0,\V c \V \sigma\right ]_-
-\frac i 2 [\hat \rho^0, \V h\cdot \V \sigma]_-
\nonumber\\
&
=-\frac 1 2 \left [\delta \hat \rho,{1\over \tau}+\V \tau^{-1} \cdot \V
  \sigma\right ]_+
-\frac 1 2 \left [\delta \hat \rho,\V d \cdot \V \sigma\right ]_-
\label{gr}
\end{align}
with the yet undetermined constants $a, \V b, \V c, \V d, \V h$.  Here the first line
describes the scalar drift which can be trivially written in anticommutators.
The commutator and anticommutator on the second line are the meanfields including a possible precession in the second and third terms. The third line expresses possible relaxations.

Now the compact equation (\ref{gr}) is decomposed into the components $\delta f$
and $\V \delta g=\V e \delta g+g\delta \V e$ with the help of (\ref{rules1}).
 The aim is to specify the three remaining vectors $\V c$, $\V d$ and $\V h$ 
such that the results of the two-fluid model (\ref{fg}) are reproduced. 
The two values $a=V_1 \delta n+V_3 \delta s$ and $\V
b=(V_2\delta n+V_4\delta s) \V e$ can be already determined since only this choice creates the meanfield terms in (\ref{fg}). It is convenient to decompose
the so far unspecified vectors 
\be
\V c&=&c \V e+c_1 \delta \V e+c_2\V e \times \delta \V e\nonumber\\
\V d&=&d \V e+d_1 \delta \V e+d_2\V e \times \delta \V e
\ee  
according to the three orthogonal directions since $\V e\cdot \partial \V e=0$ due to $|\V e|^2=1$.
We obtain from (\ref{gr}) the analogous equations to (\ref{fg})
\ba
&\partial_t \delta f+{i \V q \V p\over m_e} \delta f+e\V E \V \partial_p f^0-M_f=-{\delta
  f\over \tau}-\delta \V g \cdot  \V \tau^{-1}\nonumber\\
&+i g^0\V b\cdot q{\p p}\delta \V e
\nonumber\\
&\partial_t \delta g+{i \V q \V p\over m_e}\delta g+e\V E\V \partial_p g^0-M_g=-\delta
  f(\V e\cdot \V \tau^{-1})-{\delta g\over \tau}\nonumber\\
&
+c_1 g^0\V e\cdot (q{\p p} \V e\times \delta \V e)-c_2{q\p p}\V e\cdot \delta \V e
-i g^0 d_2(\delta \V e)^2
\label{fg1}
\end{align}
Comparing the two-fluid model (\ref{fg}) with (\ref{fg1}) we find the unique identification
$c_2=d_2=c_1=0$. Further one sees that we have to set 
\be 
{1\over \tau}={1\over \tau_+}+{2\over \tau_D};\quad \V e\cdot
\V \tau^{-1}={1\over \tau_-}
\ee
and the cross relaxation time has to be determined by
\ba
{2\over \tau_D}\delta f=-g^0(\V\tau^{-1}\cdot \delta \V e)+i g^0 \V e \cdot q\partial_p \delta \V e (V_2 \delta n+V_4 \delta s).
\label{eqe}
\end{align}
Only with these settings we obtain exactly the two-fluid model (\ref{fg}).
The resulting equation (\ref{eqe}) reveals that the cross relaxation time
$\tau_D$, see (\ref{relaxnew}),  can be
only obtained with the solution of the equation for the direction $\delta \V
e$ which, however, is a
dynamical (frequency-dependent) one. This shows what one silently
approximates when using a constant cross relaxation time $\tau_D$.

Finally from (\ref{gr}) the equation for $\delta \V e$ takes the form 
\ba
&\partial_t \delta \V e\!+\!{i \V q \V p\over m_e}\delta \V e\!+\!e\V E\V \partial_p \V e
\!+\!\V e \times \left [
c\V q\V \partial_p \V e\!+\!\V h-i(d\!-\!{\delta g\over g} d_1) \delta \V e\right ] 
\nonumber\\
&=-{\delta
  f\over g}\left [\V e\times( \V \tau^{-1}\times \V e)\right  ]\!-\!{\delta \V e\over
  \tau}\!+\!i
\V q\V \partial_p \V e (V_1\delta n\!+\!V_3 \delta s).
\label{fge}
\end{align}
The drift side has the usual form extended by a precession
term around the axes $\V e$. This spin-precession term is
expected and the values of $d$, $d_1$, $c$ and $\V h$ 
will be derived in the next section which should
include the external magnetic field as one part of the defining axes. 
The
right hand side in (\ref{fge}) contains the relaxation
mechanism which shows a coupling to the solution $\delta f$ and mean field contributions which will be obtained
from a proper microscopic theory. 

Summarizing the results of this section we have started with the often used
two-fluid model together with meanfield terms. From a proper writing of the
kinetic equation in spinor form (\ref{gr}) required by
SU(2) symmetry we have seen that a third equation for the change of the spin direction is needed. Further it reveals that the cross relaxation time can be considered only approximately as
time-independent and constant. The general possible form of the kinetic
equation for the scalar (total density) distribution, the polarization (total
spin) distribution, and the total spin direction is already settled except three
scalars $c,d,d_1$ and one vector $\V h$ to be derived from a microscopic theory, see later (\ref{constants}). It is remarkable that the demand of SU(2)
symmetry leads already to such a far leading determination of the structure of
equations.

\section{Quantum kinetic equation} 
\subsection{Green-functions}
Let us consider spin-polarized fermions which   
interact with impurity potential $V_{i}$ and are themselves covered by the Hamiltonian
\be
\hat H&=&\sum_i\Psi_i^+\left [{(\V p-e \V A(\V R_i,t))^2\over 2 m_e}+e \Phi(\V
  R_i,t) \right .\nonumber\\&&\left .-\mu_B \V \sigma \cdot \V B+\V \sigma\cdot \V
  b(\V p,\V R_i,t)+\hat V_{i}(\V R_i)\right ]\Psi_i\nonumber\\
&+&\frac 1 2 \sum_{ij}  \Psi_i^+\Psi_j^+\hat V(\V R_i-\V R_j)\Psi_j \Psi_i
\ee 
with the spinor $\Psi_i=(\psi_{i\uu},\psi_{i\dd})$ such that any of the spin-orbit couplings discussed above and the Zeeman term are included. 
We assume a two-particle interaction which has a scalar and a spin-dependent part 
\be
\hat V=V_0+\V \sigma\cdot {\V V}
\label{pot}
\ee
where the latter is responsible for spin-flip reactions.
The vector part of the potential describes e.g. spin-dependent scattering. The
scattering off impurities consists of a vector potential from magnetic impurities and a scalar one from charged or neutral impurities
\be
\hat V_i=V_{i0}+\V \sigma \cdot \V V_i.
\label{29}
\ee
In this way we have included the Kondo model as specific case which was solved exactly in equilibrium \cite{W81} and zero temperature. Here we will consider the nonequilibrium form of this model in the meanfield approximation including relaxation due to collisions. 

We use the formalism of the nonequilibrium Green's function technique in the generalized Kadanoff-Baym notation introduced by Langreth and Wilkins\cite{LW72}. The two
independent real-time correlation functions for spin-$1/2$ fermions are
defined as 
\be 
G^>_{\alpha \beta}(1,2)\!=\!\langle \psi_\alpha
(1)\psi^\dagger_\beta (2) \rangle ,\,
G^<_{\alpha
\beta}(1,2)\!=\!\langle \psi^\dag_\beta (2)\psi_\alpha (1) \rangle 
\ee
where $\psi^\dagger$($\psi$) are the creation (annihilation) operators,
$\alpha$ and $ \beta$ are spin indices, and  numbers are cumulative
variables for space and time,$1\equiv(\V r_1,t_1)$. Accordingly, all the correlation functions without
explicit spin indices,  are understood as $2\times 2$ matrices in
spin space, and they can be written in the form $\hat C=C +
\V \sigma\cdot\V C $,  where $C$ ($\V C$) is the
scalar (vectorial) part. This will result in
preservation of the quantum mechanical behavior concerning spin commutation
relations even after taking the quasi classical limits  of the kinetic
equation. The kinetic equation is obtained from the Kadanoff and Baym (KB) equation
\cite{KB62}  for the correlation function $\hat G^<$
\ba \label{KB1}
-i(\hat G_R^{-1}\!\circ\! \hat G^<\!-\!\hat G^<\!\circ\! \hat G_R^{-1})=i(\hat G^R\!\circ\! \hat \Sigma^<\!-\!\hat \Sigma^< \!\circ\! \hat G^A) 
\end{align} 
where $\hat \Sigma$ is the self-energy, and retarded
and advanced functions are defined as
\ba
C^{R,A}(1,2)&=\mp i\theta(\pm t_1\mp t_2)[C^>(1,2)+C^<(1,2)]+C^{HF}
\end{align} 
where $C^{HF}$ denotes the time-diagonal Hartree-Fock terms discussed later.
Products $\circ$ are
understood as integrations over intermediate variables, space and
time, $A\circ B=\int{d\bar t}\int{d\bar{\V r}}A(\V r_1,t_1;\bar{ \V r},\bar t)B(\bar {\V r},\bar t;\V r_2, t_2)$.  The notation of Langreth and Wilkins \cite{LW72} used here has the advantage that the correlation functions $G^\gtrless$ bear a direct physical meaning of occupation of particles and holes. Alternatively sometimes the Keldysh function is used \cite{RS86} which has to be linked to physical quantities. Moreover in this latter Keldysh-matrix notation a superfluous degree of freedom occurs canceling in any diagrammatic expansion \cite{d90} which 
is absent when directly using correlation functions and the retarded/advanced Green's functions.

We are interested in the Wigner distribution function (\ref{rhofg}) which is given by the equal time ($t_1=t_2=T,t=0$) correlation function $\hat G^<(\V k,\V R,t=0,T)$ 
\be 
\hat \rho(\V p,\V R,T)&=&\hat G^<(\V p,\V R,t=0,T)
\nonumber\\
&=&\int{d\omega\over 2 \pi} \hat G^<(\V p,\omega,\V R,T)=f+\V \sigma \cdot \V g.
\label{FG}
\ee 
We use the Wigner mixed representation in terms of the center-of-mass variables $\V R=(\V r_1+\V r_2)/2$ and the Fourier transform of the relative variables $\V r=\V r_1-\V r_2\rightarrow \V p$ which separates the fast
microscopic variations from slow macroscopic variations.

To derive the kinetic equation we expand the convolution up to second-order gradients. Matrix product terms
$A\circ B$ appearing in the KB equation can be written as 
\be 
A\circ B\rightarrow{\rm e}^{\frac i 2 (\p
\Omega \p {T}^B-\p T^A \p {\Omega}^B -\p p^A \p {R}^B + \p R^A \p {p}^B)}
AB.
\ee 
The quasi-classical limit is obtained by keeping only the first
gradients in space and time of the above gradient expansion 
\be
A\circ B\to AB+\frac i 2 \{A,B \}, 
\ee 
where curly brackets denote Poisson's
brackets, i.e., $\{A,B \}=\p \Omega A\p T B-\p T A\p \Omega B -\p p
A \p R B+ \p R A\p p B$. Therefore, in the lowest-order gradient
approximation, we have the following rule to evaluate the
commutators $[A,B]_-$ in the KB equation 
\be
[A\circ ,B]_-&\to&[A,B]_- + \frac i 2 (\{A,B\}-\{B,A\}) \nonumber \\
&=&[A,B]_- + \frac i 2 \biggl ([\p R A,\p p B]_+ - [\p p A,\p R B]_+ 
\nonumber\\&& +[\p
\Omega A,\p T B]_+ - [\p T A,\p \Omega B]_+\biggr ). 
\label{rule}
\ee
Please note that the quantum spin structure remains untouched even after gradient expansion due to the commutators.

\subsection{Gauge}
In order to prevent ambiguous results for different choice of gauges, we need to formulate the theory in  a gauge-invariant way. Under U(1) local gauge the wave function and vector potential transform
as
$
\Phi' ={\rm e}^ {-{i e\over \hbar}  \alpha (x)}\Phi
$ and
$A_\mu'= A_\mu+\partial_\mu \alpha(x)$
such that the Green-function transform itself as
\ba
G'(12)=\langle \Phi_1'\Phi_2'\rangle
&=& \langle \Phi_1\Phi_2\rangle {\rm e}^ {-{i e\over \hbar}
  \alpha(X+\frac x 2)-{i e\over \hbar}
  \alpha(X-\frac x 2)}
\nonumber\\
&=&
\langle \Phi_1\Phi_2\rangle {\rm e}^ {-{ie\over \hbar}
   x^\mu \int\limits_{-1/2}^{1/2} d\lambda \partial_\mu \alpha(X+\lambda x)}.
\end{align}
This shift can be compensated if a corresponding phase is added.
This is achieved by using a modified Fourier transform  
\be
G(kX)&=&\int d x{\rm e}^ {{i\over \hbar}x^\mu \left (k_\mu+
  e \int\limits_{-1/2}^{1/2} d\lambda A_\mu(X+\lambda x)\right )}
\nonumber\\&&\times 
G\left (X+\frac
x 2,X-\frac x 2 \right ).
 \ee
where
$A=(\phi(\V R,T),\V A(\V R,T))$ and we used four-vector notation $x=(t,\V r)$ and $X=(T,\V R)$. It is obvious that this gauge-invariant Fourier transform leads to gradients as well. To see this we consider the general gauge-invariant Fourier transform for $A_\mu=(\Phi(\V R,T), \V A(\V R,T))$ in gradient expansion
\be
\V p &=&\V k+{e} \int\limits_{-\frac 1 2}^{\frac 1 2}
d\lambda {\V A(\V R+\lambda \V r,T+\lambda \tau)}
\nonumber\\
&=&
\V k+{e} \int\limits_{-\frac 1 2}^{\frac 1 2}
d\lambda {\rm e}^ {\lambda \V r \p R^A}{\rm e}^ {\lambda \tau \partial_T^A}{\V A(\V R,T)}
\nonumber\\
&\to&\V k+{e} \int\limits_{-\frac 1 2}^{\frac 1 2}
d\lambda {\rm e}^ {-i \hbar\lambda \partial_p \p R^A}{\rm e}^ {i \hbar\lambda \partial_\omega \partial_T^A}{\V A(\V R,T)}
\nonumber\\
&=&\V k+{e} \,
{\rm sinc}{\left (\frac \hbar 2  \p p \p R^A\right )}{\rm
  sinc}{\left (\frac \hbar 2 \p \omega \p T^A \right )}{\V A(\V R,T)}
\ee
with ${\rm sinc}(x)=\sin{x}/x=1-x^2/3!+-...$ and analogously
\be
\Omega=\omega+{e} \,
{\rm sinc}{\left (\frac \hbar 2  \p p \p R^\Phi\right )}{\rm
  sinc}{\left (\frac \hbar 2 \partial_\omega \p T^\Phi \right )}{\Phi(\V R,T)}.
\ee
One sees that up to second order gradients we have correctly the following rules
for gauge invariant formulation\cite{bj91,Mo00}: (1) Fourier transform of the
difference variable $x$ to the canonical momentum $\V p$. (2)  Shift
from canonical momentum to the gauge invariant (kinematical)
momentum $k_\mu=p_\mu-e\int_{-1/2}^{1/2}{d\lambda A^\mu(X+\lambda
x)}$, which becomes $k_\mu=p_\mu-eA^\mu$ in the lowest-order
gradient expansion. (3) Then the gauge invariant functions $\bar G$
reads 
\ba 
\bar G (\V k,\omega,\V R,T)=G(\V k+e\V A,\omega +\phi ,
\V R,T)=G(\V p,\Omega,\V R,T).
\label{gauger} 
\end{align}
This treatment ensures that one has even included all orders of a constant electric field.

\subsection{Meanfield}
We want to consider now the meanfield selfenergy for impurity interactions as well as spin-orbit couplings.
A general four-point potential can be written 
\be
&&\bra {x_1x_2} \hat V \ket {x_1' x_2'}=
\sum\limits_{p,p'}{\rm e}^{-i p (x_1-x_2) +i p' (x_1'-x_2')}
\nonumber\\&&\times
\bra p \hat V\left ({x_1+x_2\over 2}-{x_1'+x_2'\over 2},{{x_1+x_2\over 2}+{x_1'+x_2'\over 2}\over 2}\right )\ket {p'}\nonumber\\
&&=\hat V_{-p,p'}\delta\left ({x_1+x_2\over 2}-{x_1'+x_2'\over 2}\right ) 
\ee 
with
\be
\hat V_{-p,p'}=\left \{ \begin{matrix}V_0(p'-p)\cr \V \sigma \cdot \V V(p'-p)\cr {i \lambda^2\over \hbar} \V \sigma \cdot (\V p\times \V p') V(p'-p)\end{matrix}\right .
\label{mf0}
\ee
for scalar, magnetic impurities (\ref{29}), the two-particle interaction (\ref{pot}), and extrinsic spin-orbit coupling (\ref{100}). 
Since the potential is time-local, the Hartree-meanfield is the convolution with the Wigner function (\ref{FG}) and written with Fourier transform of difference coordinates
\be
\hat \SigmaĤ(p,R,T)=\sum\limits_{R'qQ}{\rm e}^{i q (R-R')}\hat \rho(Q+p,R',T) \hat V_{{q-Q\over 2},{q+Q\over 2}}.
\ee
Due to the occurring product of the potentials (\ref{pot}), (\ref{29}) and the Wigner function
(\ref{FG}) one has
\be
\hat V \hat \rho=V_0 f+{\V g}\cdot {\V V}+{\V \sigma}\cdot [f {\V V}+V_0 {\V g}+i ({\V V}\times {\V g})].
\label{mf2}
\ee
The last term in (\ref{mf2}) is absent since we work in symmetrized products as they appear on the left side of (\ref{KB1}) from now on.
Consequently the selfenergy possesses a scalar and a vector component
\be
\hat \Sigma^{H}(\V p,\V R,T)=\Sigma_0(\V p,\V R,T)+\V \sigma\cdot {\V \Sigma}^{H}(\V p,\V R,T).
\label{selfe}
\ee

The interaction between a conduction electron and the magnetic
impurity $\V \sigma \cdot \V V_i$ where the direction of $\V V_i$ is the local magnetic
field deserves some more discussion. We assume that this magnetic field is randomly distributed on different
sites within an angle $\theta_l$ from the $\V e_z$ direction. The directional
average \cite{CH98} leads then to 
\be
\overline{\sum\limits_p f \V V}=|V|{\sin \theta_l\over \theta_l} \V e_z n=\V
V(q) n
\nonumber\\
\overline{\sum\limits_p \V g \V V}=|V|{\sin \theta_l\over \theta_l} \V e_z 
s=\V V(q) s.
\label{direct}
\ee 
The angle $\theta_l$ allows us to describe different models. A completely random local magnetic field $\theta_l=\pi$ is used for magnetic impurities in a paramagnetic spacer layer and in a ferromagnetic layer one uses $\theta_l=\pi/4$. The latter one describes the randomly distributed orientation against the host magnet \cite{CH98}.

For impurity potentials the spatial convolution with 
the density and spin polarization reads when Fourier transformed, $\V
R\to \V q$,
\be
\Sigma^{\rm imp}_0(p,q,T)&=&
n(q) V_0(q)+\V s(q)\cdot \V V(q)
\nonumber\\
\V \Sigma^{\rm imp}(p,q,T)&=&
\V s(q) V_0(q)+n(q) \V V(q)
.
\label{mf1}
\ee
For extrinsic spin-orbit coupling we obtain 
\ba
\Sigma^{\rm s.o.}_0(\V p,\V q,T)=
&i{\lambda^2\over \hbar^2} V(q)\left [m_e (\V d_{j}(q)\times \V q)_j-\V s(q) \cdot (\V p\times \V q)\right ]
\nonumber\\
\V \Sigma^{\rm s.o.}(\V p,\V q,T)=
&i{\lambda^2\over \hbar^2} V(q)\left [m_e (\V j(q)\times \V q)-n(q) (\V p\times \V q)\right ].
\label{ss0}
\end{align}
The used particle density and current are 
\be
n=\sum\limits_p  g(\V p,\V q,T);\quad \V j=\sum\limits_p {\V p\over m_e} \, g(\V p,\V q,T)
\label{nj}
\ee
and the spin polarization and spin current
\ba
\V s=\sum\limits_p  \V g(\V p,\V q,T);\quad \V d_{i}=\sum\limits_p {\V p \over m_e}\, [\V g(\V p,\V q,T)]_i;\quad d_{ij}=S_{ji}
.
\label{sj}
\end{align}
Please note the summation over indices in the first line of (\ref{ss0}) after cross products.


Collecting these results the inverse retarded Green's function reads
\be
\hat  G_R^{-1}(\V k,\Omega,\V R,T)=\Omega-H-\V \sigma \cdot {\V {\Sigma}}(\V k,\V R,T)
\label{gr1}
\ee
with the effective scalar Hamiltonian
\be
H={k^2\over 2 m}+\Sigma_0(\V k,\V R,T)+e \Phi(\V R, T)
\ee
and ${\V k}={\V p}-e {\V A}({\V R},t)$. We have summarized the Zeemann term, the intrinsic spin-orbit coupling, and the vector part of the Hartree-Fock selfenergy due to impurities and extrinsic spin-orbit coupling into an effective selfenergy 
\be
\V {\Sigma}={\V \Sigma}^{H}(\V k,\V R,T)+\V b(\V k,\V R,T)+\mu_B \V B
\label{sig}
\ee
such that the effective Hamiltonian possesses Pauli structure
\be
\hat H_{\rm eff}=H+\V \sigma\cdot \V \Sigma.
\ee
Please note that one can consider (\ref{sig}) as an effective Zeeman term where the spin-orbit competes with the magnetic field leading to additional degeneracies in Landau levels \cite{Shen:2004}.

\subsection{Commutators}
Now we are ready to evaluate the commutators according to (\ref{rule}). In the
following we drop the vector notation where it is obvious. We calculate first the commutator with the scalar parts of (\ref{gr1}) where we use the gauge-invariant Green's function (\ref{gauger})
such that one has 
\be
\partial_p G&=&\partial_k \bar G\nonumber\\
\partial_R G&=&\partial_R \bar G-e\nabla \Phi\partial_\omega \bar G-e(\nabla
A_i)\partial_{k_i}\bar G\nonumber\\
\p T G&=&\p T \bar G-e \p T A \p k \bar G-e \p T \Phi \p \omega \bar G\nonumber\\
\p \Omega G&=&\p \omega \bar G.
\label{gaugerule}
\ee
Further one calculates
\ba
&\partial_T H=e\dot{\Phi}-e {k\over m_e} \dot{A}-e \dot A \p k \Sigma_0+\dot \Sigma_0
\nonumber\\
&\partial_p H={k\over m_e}+\p k \Sigma_0
\nonumber\\
&\partial_R H=e \partial_R \Phi\!+\!{1\over m_e} \left [k\!\times\! \partial_R\!\times\!
    (p\!-\!eA)\!+\!(k\cdot\partial_R)(p\!-\!eA)\right ]
\nonumber\\&
\qquad\quad+\p R \Sigma_0 -e\p R A_i\p {k_i}\Sigma_0\nonumber\\
&=e\nabla \Phi\!-\!{k\over m_e}\times e B \!-\!{e\over m_e}(k\cdot \nabla) A\!+\!\p R \Sigma_0 \!-\!e\p R A_i\p {k_i}\Sigma_0
\end{align}
where we have used $\frac 1 2 \nabla u^2=u\times\nabla\times
u+(u\cdot\nabla)u$. We obtain for the commutator with the scalar part of (\ref{gr1}) according to (\ref{rule}) 
\begin{align}
&\frac 1 i[(\Omega-H)\circ ,G^<]_-\to
\biggl [\partial_T +(\dot \Sigma_0+e v E)\partial_\omega
\nonumber\\&
+v \partial_R +(e E+e v\times
B-\p R\Sigma_0)\partial_k \biggr ] \bar G^<
\label{kin1}
\end{align}
where the mean velocity of the particles is given by
\be
v={k\over m_e}+\p k \Sigma_0
\label{vel}
\ee
and $E=-e\p R \Phi-e\dot A$ and $B=\p R \times A$.

In order to get the equation for the Wigner distribution we integrate over frequency 
and the second term on the right hand side of (\ref{kin1}) disappears. This term has the structure of the power supplied to the particles which is composed of the contribution by the electric field and the time change of the scalar field which feeds energy to the system. The first and third part of (\ref{kin1}) together are the co-moving time derivative of a particle with velocity (\ref{vel}). The fourth term in front of the momentum derivative of $\bar G^<$ represents the forces exercised on the particles which appears as the Lorentz force and the negative gradient of the scalar part of the selfenergy which acts therefore like a potential.

Next we calculate the commutator (\ref{rule}) with the vector components of (\ref{gr1}). Therefore we employ the relations
\be
[\V \sigma \cdot \V A,\hat  G^<]_+&=& 2 \V \sigma \cdot \V A \,G^<+2 \V A \cdot \V G^<
\nonumber\\
\left [\V \sigma \cdot \V A,\hat  G^<\right ]_-&=& 2 i \V \sigma \cdot (\V A \times \V G^<)
\ee
where $\hat {G}^<=G^<+\V \sigma \cdot \V G^<$
and using (\ref{gaugerule}) we obtain
\ba
&-i[-\V \sigma \cdot \V \Sigma \circ, \hat G^<]_-\to
\biggl [
\partial_T +(\dot \Sigma_0+e v E)\partial_\omega+ v \partial_R
\nonumber\\
&
+(e E+e v\times B-\p R\Sigma_0)\partial_k 
 \biggr ] 
{\V {\bar G}}^<-2 (\V \Sigma \times {\V {\bar G}}^<)\nonumber\\
&+\left [
(\dot {\V \Sigma}+e v E )\partial_\omega+\partial_k \V \Sigma \partial_R
+(e \partial_k \V \Sigma\times B-\p R \V  \Sigma)\partial_k 
\right ] 
{\bar G}^<.
\nonumber\\
&
\label{kin2}
\end{align}
We recognize the same drift terms as for the scalar selfenergy components (\ref{kin1}). Additionally, the vector selfenergy couples the scalar and spinor part of the Green function by an analogous drift but controlled by the vector selfenergy instead of the scalar one.

\subsection{Coupled kinetic equations}
Integrating (\ref{kin1}) over frequency and adding (\ref{kin2}) we have the
complete kinetic equation as required from the Kadanoff and Baym equation
(\ref{KB1}). In order to make it more transparent we separate the equation
according to the occurring Pauli matrices. This is achieved by once forming
the trace and once multiplying with $\V \sigma$ and forming the trace. We obtain finally two coupled equations for the scalar and vector part of the Wigner distribution
\be 
D_T f+\V A \cdot \V g &=&0\nonumber\\
D_T \V g+\V A f
&=&2 (\V \Sigma\times \V g)
\label{kinet}
\ee
where $D_T=(\p T+\V {\cal F}\V {\p k}+\V v\V {\p R})$ describes the drift and force of the scalar and vector part with the velocity (\ref{vel}) 
and the effective Lorentz force
\be
\V {\cal F}=(e \V E +e \V v \times  \V B - \V {\p R}
\Sigma_0).
\label{lor}
\ee
The coupling between spinor parts is given by the vector drift
\be
A_i=(\V \partial_k \Sigma_i\V \partial_R-\V \partial_R\Sigma_i\V \partial_k+e(\V \partial_k\Sigma_i \times \V B)\V \partial_k).
\label{A}
\ee
Remember that we subsumed in the vector selfenergy (\ref{sig})
the magnetic impurity meanfield, the spin-orbit coupling vector, and the Zeeman term.

The term (\ref{A}) in the second parts on the left sides of (\ref{kinet}) represent the coupling between the spin parts of the Wigner distribution. The vector part contains additionally the spin-rotation term on the right hand side. 
These coupled mean field kinetic equations including the magnetic and electric field, Zeeman coupling, and spin-orbit coupling are the final result of the section.
On the right hand side one has to consider additionally collision integrals which can be derived from the KB equation taking the selfenergy beyond the meanfield approximation. In the simplest way we will add a relaxation time. 

The system (\ref{kinet}) is the main result of this paper and the basis for the further discussion including collective modes in the second paper of the series. Therefore it is time now to compare with other approaches and lay out the generalizations obtained here. If one neglects the coupling of the scalar distribution $f$ to the vector distribution $\V g$ in the second equation of (\ref{kinet}) one has the Eilenberger equation \cite{SDGR06} extended here by magnetic and electric fields as well as selfenergy effects. Compared to \cite{Me89} we write both scalar and vector components and have included meanfield quasiparticle renormalizations and the vector self energy. The coupling of the vector equation to the scalar one has been neglected in \cite{LeRi68,Mi05} too but selfconsistent quasiparticle energies and Zeeman fields have been taken into account. In \cite{MeSt95,GoRu95} only the transverse components have been considered \cite{MeSt95} which approximates two of the four degrees of freedom in (\ref{kinet}). The same reduction of degrees of freedom at the beginning has been used by the projection technique in \cite{NaTa91,NTL91} since the focus of all these papers had been on the proper collision integral instead. Selfconsistent quasiparticle equations have been presented in \cite{JeMu89,Bas00} which have been decoupled by Landau Fermi-liquid assumptions \cite{MeMu92} and variational approaches. The coupled kinetic equations have been derived without spin-orbit coupling terms and reduced vector selfenergies in \cite{WNCC02} which disentangle the equation of moments. Here we present all these effects without the assumptions found in different places of the above-mentioned approaches.

\subsection{Quasi stationary solution}

The time-independent stationary solution should obey the stationary mean-field equation (\ref{kinet}) since any collision integral is then zero providing the Fermi distribution. However, the arguments and functional dependence as well as spin structure of the solution are already determined from the stationary equation of the mean field equation (\ref{kinet}). We write them in formal notation
\be
D f+\V A \cdot \V g&=&0\nonumber\\
D\V g+\V A f&=& 2 (\V \Sigma \times \V g)
\label{statio}
\ee   
with
\be
D=\{\epsilon_k,...\},\quad A_i=\{\Sigma_i,...\}
\label{DA}
\ee
and the Poisson bracket $\{a,b\}=\V\partial_k a\cdot \V \partial_R b-\V \partial_R a\cdot \V \partial_kb$.
The electric field is given by a scalar potential $e\V E(R)=-\V \nabla \Phi(R)$ such that we have the quasiparticle energy
\be
\epsilon_k(R)={k^2\over 2 m_e}+\Sigma_0(k,R)+\Phi(R).
\ee
The choice of gauge is arbitrary since we have ensured that the kinetic equation is gauge invariant.

We rewrite (\ref{statio}) into one equation again by the spinor representation $\hat \rho=f+\V \sigma\cdot \V g$ using the identity 
\ba
&\V c\cdot \V g+ (\V \sigma \cdot \V c) f-2 \V \sigma \cdot (\V \sigma\times \V g)=
\nonumber\\&
\V \sigma \cdot {\V c+2 i\V \sigma\over 2} \hat \rho+\hat \rho \V \sigma \cdot {\V c-2 i\V
  \sigma\over 2}
\label{identity}
\end{align}
to arrive at
\be
[D+\V \sigma \cdot \V A,\hat \rho]_++2 i [\V \sigma\cdot \V \Sigma,\hat \rho]_-=0
\label{Feq}
\ee 
which is equivalent to (\ref{statio}).
Now we search for a solution which renders both anticommutator and commutator 
zero separately. 

For the anticommutator, the equation 
\be
(D+\V \sigma \cdot \V A) \hat \rho=0 
\ee
and correspondingly $\hat \rho(D+\V \sigma \cdot \V A) =0$ are solved by any function of the argument
\be
\hat \rho_0 \left [\epsilon_k(R)+ \V \Sigma(k,R)\cdot\V \sigma \right ]
\ee
due to (\ref{DA}).
Employing the relation
\ba
{\rm e}^{\V \sigma\cdot \V \Sigma}&=&\cosh{|\V \Sigma|}+\V \sigma\cdot \V e\, \sinh{|\V \Sigma|}=\sum_{s=\pm} \hat P_s {\rm e}^{s |\V \Sigma|}
\end{align}
with the projectors $\hat P_\pm=\frac 1 2 (1\pm \V e\cdot \V \sigma)$ and $\V e={\V \Sigma/ |\V \Sigma|}$
we have to have the stationary solution in the form
\be
\hat \rho_0\left [\epsilon_k(R)+ \V \Sigma(k,R)\cdot\V \sigma \right ]=\sum_{\pm} P_\pm \hat \rho_\pm(\epsilon_k\pm |\V \Sigma|).
\label{72}
\ee

The demand of vanishing commutator in (\ref{Feq}) works further down the still general possibility of distribution $\hat \rho_\pm=\bar f_\pm+\V \sigma\cdot \V g_\pm$. In fact it demands
\be
0=[\V \sigma\cdot \V \Sigma,\hat \rho]_-=[\V \sigma\cdot \V \Sigma,\V \sigma\cdot \V {g}]_-=i\V \sigma\cdot (\V \Sigma \times \V {g})
\ee
which implies that $\V {g}=\V e g$ with the effective direction $\V e=\V \Sigma/|\V \Sigma|$. Together with (\ref{72})
we obtain the stationary solution of (\ref{Feq}) and consequently of (\ref{statio}) to have the form
\be
\hat \rho({\hat \varepsilon})=\sum\limits_{\pm}\hat P_\pm f_\pm
&=&{f_++f_-\over 2}+\V \sigma\cdot \V e \,\,{  f_+-f_-\over 2}\nonumber\\
&\equiv&\rho+\V \sigma \cdot \V \rho
\label{solF}
\ee
with $f_\pm=\bar f_\pm+g_\pm=f_0(\epsilon_k(R)\pm|\V \Sigma(k,R)|)$ and $f_0$ an unknown scalar function which is determined by the vanishing of the collision integral to be the Fermi-Dirac distribution.

\subsection{Currents\label{chap_current}}

Due to the spin-orbit coupling (\ref{so}) the current possesses an normal and anomaly part. Using $[\V b(\V p),\V x]=-i \hbar \partial_{\V p} \V b(\V p)$ from elementary quantum mechanics we have 
\be
\hat v_j=\frac i \hbar [\hat H,\hat {x}_j]
=v_j+ \partial_{p_j} \V b \cdot \V \sigma
\ee
and the quasiparticle velocity $v_j=\p {p_j} \epsilon$ if the single particle Hamiltonian is given by the quasiparticle energy $\epsilon(p)$.
Together with the Wigner function $\hat \rho=f+\V g\cdot \V \sigma$ 
one has
\be
\hat \rho \hat v_j=f v_{i}+\V g \cdot \V \beta_j+\V \sigma\cdot (v_{j}\V g+f \partial_{p_j} \V b +i \partial_{p_j} \V b \times \V g)
\ee
and the particle current and spin current density reads
\ba
\hat J_j=\frac 1 2 \sum\limits_p [\hat \rho, v_j]_+
&=\sum\limits_p\left [
f v_{j}\!+\!\V g \cdot \partial_{p_j} \V b \!+\!\V \sigma\cdot (v_{j}\V g\!+\!f \partial_{p_j} \V b )
\right  ]\nonumber\\
&=J_j+\V \sigma\cdot \V S_j.
\label{current}
\end{align}
The scalar part describes the particle current $\V J=\V J^n+\V J^a$ consisting of a normal and anomaly current and the vector part describes the spin current $S_{ij}$ not to be cinfused with the polarization $\V s$. 

The stationary solution allows one to learn about the seemingly cumbersome structure of the particle current (\ref{current}) consisting of normal and anomalous parts. In fact both parts are necessary to guarantee the absence of particle currents in stationary spin-orbit coupled systems. In fact both parts separately are nonzero and only their sum vanishes. We expand the normal particle current with (\ref{solF}) and $\V \Sigma=\V \Sigma_n+\V b$ linear in the spin-orbit coupling $\V b$ to get
\ba
&J^n_i=\frac 1 2 \sum\limits_p \p {p_i} \epsilon \left [ f(\epsilon_p+\Sigma)+f(\epsilon_p-\Sigma)\right ]
\nonumber\\
&=\frac 1 2 \sum\limits_p \p {p_i} \epsilon {\V \Sigma_n\cdot \V b\over
\Sigma_n}\p \epsilon\left [
f(\epsilon_p+\Sigma_n)-f(\epsilon_p-\Sigma_n)\right ].
\end{align}
The anomalous current reads linear in $\V b$
\be
J^a_i&=&\frac 1 2 \sum\limits_p (\p {p_i} \V b)\cdot \V e \left [ f(\epsilon_p+\Sigma)-f(\epsilon_p-\Sigma)\right ]
\nonumber\\
&=&
\frac 1 2 \sum\limits_p {\V \Sigma_n\cdot \p {p_i} \V b\over
\Sigma_n}\left [
f(\epsilon_p+\Sigma)-f(\epsilon_p-\Sigma)\right ].
\ee
Combining both currents one obtains
\be
J_i&=&J^a_i+J^n_i
\nonumber\\&=&
{\V \Sigma_n\over \Sigma_n}\cdot \frac 1 2 \sum\limits_p\p {p_i} \left \{\V b \left [f(\epsilon_p+\Sigma)-f(\epsilon_p-\Sigma)\right]\right \}
\nonumber\\&=&0
\ee
as one should. This demonstrates the importance of the anomalous current. One can consider the spin-orbit coupling as a continuous current of normal quasiparticles compensated by the spin-induced one. Any disturbance and linear response will lead to interesting effects due to this disturbed balance such as the anomalous Hall and spin-Hall effects discussed in Sec. IV.

\subsection{Selfconsistent precession direction}

The meanfield approximation establishes a nonlinear relation for parameters of the distribution functions. The quasi-stationary distribution (\ref{solF}) is determined by the selfenergy (\ref{mf1}), (\ref{mf2}) and (\ref{sig}) which in turn is determined again by the distribution. Without spin polarization usually this leads to the selfconsistent determination of the chemical potential.

Now we have to accept that the spin precession direction obeys a similar selfconsistency and has to be determined accordingly. 

Let us assume that the external magnetic field is in the z-direction as is the mean local magnetic field of the magnetic impurities (\ref{direct}) and write $\mu_B B^{\rm eff}=n V+\mu_B B$. Since the effective local spin (precession) direction is $\V e=\V \Sigma/\Sigma$ the mean polarization reads with (\ref{sig})
\be
\V s&=&\sum\limits_p {\V b(p)+\mu_B B^{\rm eff} \V e_z+V_0 \V s\over |\V \Sigma|} g
\nonumber\\
&=&{\sum\limits_p \V b{g\over |\V \Sigma|}\over 1-V_0\sum\limits_p {g\over |\V \Sigma|}}+\mu_B B^{\rm eff} \V e_z {\sum\limits_p {g\over |\V \Sigma|}\over 1-V_0\sum\limits_p {g\over |\V \Sigma|}}.
\ee
Since $|\V \Sigma|=|\V b(p)+\mu_B B^{\rm eff} \V e_z+V_0 \V s|$ one recognizes the selfconsistent equation for the mean polarization $\V s$ which in turn determines the local spin precession direction $\V e$. 
In other words the usual selfconsistency due to the meanfields is extended towards a scalar density and a vector polarization yielding the values and the selfconsistent precession direction. The procedure is as follows. One calculates the density and spin-polarization according to (\ref{quasin}) where the distributions (\ref{solF}) are dependent on the vector selfenergy (\ref{sig}) which in turn is again determined by the density and spin polarization. The scalar selfenergy we absorb into an effective chemical potential.
This selfconsistent precession direction is solely due to the momentum dependence of the spin-orbit coupling $\V b$. 

Let us inspect all the equations in first order of spin-orbit coupling. We write $\V \Sigma=\V \Sigma_n +\V b_p$ where we denote the momentum-independent selfenergy with $\V \Sigma_n=n\V V+V_0 \V s+\mu_B \V B$. 
We expand all directions in first order of $\V b$.

The direction of effective polarization becomes
\be
\V e &=& {\V \Sigma\over |\Sigma|}=\V e_z \left ( 1-{b_\perp^2\over 2}\right )+{\V b_\perp}\left (1-{b_3}\right )
\label{guess}
\ee
where we will use the convenient separation in the
z-direction and the perpendicular direction
\be
{\V b_p\over \Sigma_n}=\V b_\perp+\V e_z b_3.
\ee

The first impression of (\ref{guess}) suggests that one has a deviation from the z-direction due to the perpendicular direction $\V b_\perp$. Let us calculate the selfconsistency and see what remains from this deviation.
Since the distribution functions in equilibrium are functions of $|\V \Sigma|$ according to (\ref{solF}), i.e. a function of $b_\perp^2$ and $b_3$, and since the latter ones are even in the momentum direction, the distributions are even in the momentum direction. Therefore the polarization becomes
\be
\V s=\sum\limits_p g\V e =\V e_z \sum\limits_p g\left (1-{b_\perp^2\over
    2}\right )=\V e_z \left (s_0-{B_g^2\over 2}\right )
\label{s}
\ee
with
\be
s_0&=&\sum\limits_p g;\qquad B_g^2=\sum\limits_p b_\perp^2 g.
\label{Bg}
\ee
Now the effective precession direction $\V e=\V \Sigma/|\V \Sigma|$ is
seen to be in the z-direction up to second order in the spin-orbit coupling in contrast to our first view (\ref{guess}).

It is instructive to calculate the selfconsistent precession explicitly up to any order now for zero temperature in quasi two dimensions and linear Dresselhaus $\beta=\beta_D$ or Rashba $\beta=\beta_R$ spin-orbit coupling. We have the density and the polarization
\be
n&=&\sum\limits_p f= {m_e\over 2 \pi \hbar^2} \left ( \epsilon_f+\epsilon_\beta\right )
\nonumber\\
s&=&\sum\limits_p g=-{m_e\over 2\pi \hbar^2} \sqrt{\epsilon_\beta(\epsilon_\beta+2 \epsilon_f)+\Sigma_n^2}
\label{ns}
\ee 
with the spin-orbit energy $\epsilon_\beta=m_e\beta^2$. On sees how the Fermi energy $\epsilon_f$ is shifted by the spin-orbit coupling. The effective Zeeman term $\Sigma_n$ determines the polarization in the absence of spin-orbit coupling as $s/n=-\Sigma_n/\epsilon_f$. Since $\Sigma_n=\mu_B B^{\rm eff}+ s V_0$ with $\mu_B B^{\rm eff}=n V+\mu_B B$ we might conclude from the quadratic equation for $s$ in (\ref{ns}) that the selfconsistent polarization becomes
\ba
&s^{\rm self}\nonumber\\
&={m_e\over 2 \pi\hbar^2} {{m_e V_0\over 2 \pi \hbar^2} \mu_B B^{\rm eff} \!\pm\! \sqrt{(\mu_B B^{\rm eff})^2\!+\!\epsilon_\beta (\epsilon_\beta\!+\!2 \epsilon_f) \left ({m_e V_0\over 2 \pi \hbar^2}\right )^2}\over 1-\left ({m_e V_0\over 2 \pi \hbar^2}\right )^2}
\label{scons}
\end{align}
However, this procedure oversees just the selfconsistent precession $\V e ={\V \Sigma/ |\V \Sigma|}$. In fact instead of (\ref{ns}) we have to calculate the vector quantity
\be
\V s&=&\sum\limits_p \V e g=\int\limits_0^\infty {dp p\over (2 \pi \hbar)^2} \int\limits_0^{2\pi}\V e g
\nonumber\\
&=&\V e_z \int\limits_{p_1}^{p_2} {dp p\over 4 \pi \hbar^2} {\mu_B B^{\rm eff}+V_0 s_z\over \sqrt{\beta^2 p^2+(\mu_B B^{\rm eff}+V_0 s_z)^2}}
\nonumber\\
&=&
-\V e_z {m_e\over 2\pi \hbar^2} \left (\mu_B B^{\rm eff} +V_0 s_z\right )
\label{vs}
\ee
with 
\ba
{p_{1/2}^2\over 2 m_e}=\epsilon_f\!+\!\epsilon_\beta\pm \sqrt{\epsilon_\beta (2 \epsilon_f\!+\!\epsilon_\beta)\!+\!(\mu_B B^{\rm eff}\!+\!V_0 s_z)^2}
\end{align}
originating from the zero-temperature Fermi functions. We obtain just the result (\ref{scons}) but without spin-orbit coupling $\epsilon_\beta\to 0$ 
\be
\V s^{\,\,\rm self}=-\V e_z {m_e\over 2 \pi \hbar^2} {n V+\mu_B B\over 1+{m_e V_0\over 2 \pi \hbar^2}}
\label{selfs}
\ee
which is quite astonishing. Though the selfconsistent Fermi functions and the selfconsistent precession both contain an involved spin-orbit coupling separately, they cancel each other in the polarization in quasi two dimensions and for Rashba or Dresselhaus coupling. 

It remains to show that this observation is consistent with the general expression for the linearized result (\ref{s}). With the help of (\ref{ns}) one obtains in fact just (\ref{selfs})
\ba
&\V s=\V e_z \left (\sum\limits_p g\!-\!\frac 1 2 \sum\limits_p g b_\perp^2 \right )
\nonumber\\
&=-{m_e\V e_z\over 2 \pi \hbar^2} \left ( \mu_B B^{\rm eff}\!+\!V_0 s_z \right )\!+\!o(\beta^3)
=- {m_e\V e_z\over 2 \pi \hbar^2} {n V+\mu_B B\over 1+{m_e V_0\over 2 \pi \hbar^2}}
\end{align}
and the effective magnetic field becomes renormalized 
\be
\mu_B B^{\rm eff}={n V+\mu_B B}\to{n V+\mu_B B\over 1+{m_e V_0\over 2 \pi \hbar^2}}
\ee
due to selfconsistency.
We can conclude that the selfconsistency will determine an effective precession direction deviating from the direction of the external magnetic field due to the spin-orbit coupling. However this effect is of higher than second order in spin-orbit coupling $\V b$ and vanishes in quasi two-dimensional systems at zero temperature and linear spin-orbit coupling.

\section{Balance equation} 

\subsection{Linearization to external electric field}
We consider now the linearization of kinetic equation (\ref{kinet}) with respect to an external electric field, no magnetic field, and a homogeneous situation. We Fourier transform the time $\partial_t\to -i\omega$ and the spatial coordinates $\V \partial_R\to i \V q$. 
The distribution is linearized according to $\hat \rho(pRT)=f(p)+\delta f(pRT)+\V \sigma \cdot
[\V g(p)+\delta \V g(pRT)]$ due to the external electric field perturbation
$e \delta \V E=e \V E(R,T)=-\nabla \Phi$. 
Further we assume a collision integral of the relaxation time approximation \cite{HaMo92}
\be
-\frac 1 2 [\hat \tau^{-1}, \delta \hat \rho^l]_+
\ee
with a vector and scalar part of relaxation times
$\hat \tau^{-1}=\tau^{-1}+\V \sigma \cdot \V \tau^{-1}$ and
\be
\tau={\tau^{-1}\over \tau^{-2}-|\V \tau^{-1}|^2},\quad \V \tau=-{\V \tau^{-1}\over \tau^{-2}-|\V \tau^{-1}|^2}.
\ee

The scalar relaxation is assumed not towards the absolute equilibrium $f_0(\epsilon \pm |\Sigma|-\mu)$ characterized by the chemical potential $\mu$ but towards a local one $f^l=f_0(\epsilon \pm |\Sigma|-\mu-\delta \mu)$. The latter one can be specified such 
\be
\delta n&=&\sum\limits_p (f-f_0)=\sum\limits_p (f-f^l+f^l-f_0)
\nonumber\\
&=&\sum\limits_p (f^l-f_0)=\partial_\mu n \,\delta \mu
\ee
such that the density is conserved \cite{Mer70,D75}
as expressed in the step to the second line.
Therefore the relaxation term becomes
\be
-{\delta \hat \rho^l\over \tau}=-{\delta \hat \rho\over \tau}+{\delta n\over \tau\partial_\mu n}\partial_\mu \hat \rho_0.
\ee
In this way the density is conserved in the response function which could be extended to included more conservation laws \cite{Ms01,Ms12}. If we consider only density conservation but no polarization conservation, of course, we can restrict ourselves to the $\p \mu f_0$ terms.

Abbreviating $-i \omega+i \V p\cdot \V q/m+ \tau^{-1}=a$ and $iq\p p\V \Sigma+\V \tau^{-1}=\V b$, 
the coupled kinetic equations (\ref{kinet}) take the form
\be
a \delta f+\V b \delta \V g&=&S_0\nonumber\\
a \delta \V g+\V b \delta f-2 \V \Sigma\times \delta \V g&=&\V S
\label{kineq}
\ee
with $e\V E=-i \V q \Phi$ and
\ba
S_0&=iq\p p f (\Phi+\delta \Sigma_0)+i q \p p \V g \cdot \delta \V \Sigma+{\delta n\over \tau\partial_\mu n}\partial_\mu f_0\nonumber\\
{\V S}&=iq\p p \V g (\Phi+\delta \Sigma_0)+i q\p p f \delta \V \Sigma+2 (\delta \V \Sigma \times \V g)+{\delta n \partial_\mu \V g\over \tau\partial_\mu n}.
\label{source}
\end{align}
In order to facilitate the vector notation we want to understand $q\p p=\V q\cdot \V{\p p}$ in the following.

\subsection{Density and spin current}

The linearized kinetic equations (\ref{kineq}) allow us to write the balance equations for the magnetization density $\delta \V s$, the density $\delta n$,  and the currents by integrating over the corresponding moments of momentum
\ba
\p t \delta n+\p {R_i} \V J_i+\V \tau^{-1}\cdot \delta \V s&=0
\nonumber\\
\p t \delta \V s+\p {R_i} \V S_i+\V \tau^{-1} \delta n-2\sum\limits_p\V \Sigma \times \delta \V g &=2\delta \V \Sigma \times \V s
\label{balancens}
\end{align}
where we Fourier transformed the wavevector $q$ back to spatial coordinates $R$. Exactly the expected density currents and magnetization currents (\ref{current}) appear
\be
\V J&=&\sum\limits_p \left (\p p \epsilon_p \delta f+\p p \V b\cdot \delta \V g\right )\nonumber\\
\V S_i&=&\sum\limits_p \left (\p {p_i} \epsilon_p \delta \V g+\p {p_i} \V b \delta f \right ).
\label{currents}
\ee
The right hand side of (\ref{balancens}) can be reshuffled to the left since $\delta \V \Sigma=V_0\delta \V s+\V V \delta n$. The only problem makes the term
$\sum\limits_p \V \Sigma \times \delta \V g$ since the momentum-dependence of the spin-orbit coupling prevents the balance equations from being closed. We need the complete solution of $\delta \V g$ in order to write the correct balance equation for the magnetization density. This will be given in the second part of this paper series.

The method of moments does not yield a closed system of equations since the density couples to the currents, the balance for the currents to the energy currents and so on. Only with specific approximations these equations can be closed. One can find a great variety of methods in the literature. Many treatments neglect certain Landau-liquid parameters \cite{Mi04} based on the work of \cite{Le70}. A more advanced closing procedure was provided by \cite{Mi05} where the energy dependence of $\delta \V s$ was assumed to be factorized from space and direction $\V p$ dependencies.

We will not follow these approximations but solve the linearized equation exactly in the second part of this paper series to provide the solution of the balance equations and the dispersion exactly. Amazingly this yields quite involved and extensive structures with many more terms then usually presented in the literature.

\subsection{Comparison with two-fluid model}
We are now in a position to compare the result of microscopic theory with the two-fluid model (\ref{fg}) and the form for the direction (\ref{fge}) extracted from general symmetric considerations. The first observation is that the momentum-dependence of the direction $\V e$ due to, e.g., spin-orbit coupling is not covered by the original two-fluid model. We can, however, redefine certain relaxation times to account for these effects as we did in (\ref{eqe}). The price to pay was that one has to consider a third equation for the precession direction. The undetermined constants in (\ref{fge}) are now derived to be 
\be
\V h=2 \delta\V \Sigma, \qquad d=-2 i \Sigma,\qquad  d_1=c=0
\label{constants}
\ee 
which one can see from decomposing (\ref{kineq}) and (\ref{source}) into equations for $\delta f$, $\delta g$ and $\delta \V e$ and compare with (\ref{fg1}) and (\ref{fge}). 

The problem with the two-fluid model becomes apparent if we try to extract the values for the meanfields (\ref{potmean}). One obtains the unique identification
\be
V_1 \delta n+V_3\delta s=\delta \Sigma_0
\label{V13}
\ee
and two different forms from the $\delta f$ and the $\delta g$ equation
\be
\V e \cdot q\partial_p \V \Sigma \delta g
+\eta
&=&q\partial_p g (V_2\delta n+V_4\delta s-\V e \cdot \delta \V \Sigma)
\nonumber\\
\V e \cdot q\partial_p \V \Sigma \delta f&=&q\partial_p f (V_2\delta n+V_4\delta s-\V e \cdot \delta \V \Sigma)
\ee
with $\eta=(\Sigma\delta \V e-\delta \V \Sigma) q\partial_p \V e g\approx 0$. One sees that we have two determining equations which are consistent only if $\delta f/q\partial_p f=\delta g/q\partial_p g$ which is not ensured in general. Therefore the mapping of the momentum-dependent spin-orbit coupling terms to a two-fluid model is not possible in general. Only if we approximate the momentum-dependence of $\V \Sigma$ by a constant direction $\V e(p)\approx\V e$ do we have the two unique equations (\ref{V13}) and
\be
V_2\delta n+V_4\delta s=\V e \cdot \delta \V \Sigma.
\label{V24}
\ee
For illustration we decompose the change of precession direction into the components proportional to the density and polarization change $\delta \V e=\V e_1 \delta n+\V e_2 \delta s$ and obtain in this way
\ba
V_1&=V_0+\V V\cdot \V e,\qquad &V_2&=&\V e\cdot \V V+V_0 \V e \cdot \V e_1,
\nonumber\\
V_4&=V_0+V_0\V e\cdot \V e_2,\qquad &V_3&=&\V e \cdot \V V+\V V\cdot \V e_2
\end{align}
which provides indeed four different meanfield potentials (\ref{potmean}) which in turn can be translated into the original $U_{ij}$ potentials. In the case that $U_{\uu\dd}=U_{\dd\uu}$ we must have $V_0 \V e\cdot \V e_1=\V V\cdot \V e_2$.

\section{Spin-Hall and anomalous Hall effect}

\subsection{Homogeneous situation without magnetic field}
It is interesting to note that the coupled kinetic equation (\ref{kinet})
allows for a finite conductivity even without collisions and a Hall effect without external magnetic field. This is due to the
interference between the two-fold splitting of the band and will be the reason for the
anomalous Hall effect. We have seen already an expression of this interference by the two compensating currents, the normal and anomalous ones, in Sec.  \ref{chap_current}.

For a homogeneous system neglecting magnetic fields we
have from (\ref{kinet})
\be
(\p t+e E\p p )f&=&0\nonumber\\
(\p t+e E\p p )\V g&=&2(\V \Sigma \times \V g).
\label{homog}
\ee
Due to the spin-precession term a nontrivial solution for the
polarization part $\V \rho$ of the distribution appears. We will solve these 
coupled equations in the following by two ways, once in the helicity basis
\cite{Liu05} and once directly in the spin basis. In order to gain trust in the result we then compare the expressions with the Kubo formula.

With the help of an effective Hamiltonian
\be
H=\epsilon+\V \sigma \times \V \Sigma
\ee
with $\epsilon={p^ 2\over 2 m}+\Sigma_0+e \Phi$ and the spin-orbit coupling as
well as the meanfield by magnetic impurities summarized in $\V\Sigma$, one can rewrite 
both coupled kinetic equation $\hat {\bar \rho}=\rho+\V \sigma \times \V \rho$ into
\be
(\p t+e E\p p ) \hat {\bar \rho}+i [H,\hat {\bar \rho}]_-=0.
\label{hel1}
\ee

\subsection{Helicity basis}
Now we go into the helicity basis which means we use the eigenstates 
\be
H\ket \pm =\epsilon^\pm \ket \pm
\ee
and using the convenient notation 
\be
\Sigma_x-i \Sigma_y=\Sigma{\rm e}^{-i\varphi}:\quad |\Sigma|=\sqrt{\Sigma_x^2+\Sigma_y^2+\Sigma_z^2}
\label{polar}
\ee
we have
\be
\epsilon^\pm&=&\epsilon\pm |\Sigma|\nonumber\\
|\Sigma|^2&=&\Sigma^2+\Sigma_z^2\nonumber\\
\ket \pm&=&{1\over \sqrt{2|\Sigma|}}\begin{pmatrix}-{\rm e}^{-i \varphi}\sqrt{|\Sigma|\pm \Sigma_z}\cr \mp \sqrt{|\Sigma|\mp \Sigma_z}\end{pmatrix}.
\label{pm}
\ee
The transformation matrix is $U=(\ket +,\ket -)$ and the Hamiltonian becomes
diagonal
\be
\bar H=U^+HU=\begin{pmatrix}\epsilon^+&0\cr 0&\epsilon^-\end{pmatrix}
\label{hbar}
\ee
and since the transformed spin-projection operators read
\be
\bar P_+=\begin{pmatrix}1&0\cr 0&0\end{pmatrix},\quad \bar
P_-=\begin{pmatrix}0&0\cr 0&1\end{pmatrix}
\ee
the equilibrium distribution (\ref{solF}) becomes diagonal
\be
\hat {\bar \rho}=\sum\limits_{i=\pm}\hat P_i f_i=\begin{pmatrix}f_+&0\cr 0&f_-\end{pmatrix}
\ee
or in general
\be
\hat {\bar \rho}=\bar \rho+\V {\bar \rho} \cdot U^+\V \sigma U.
\label{F0}
\ee
 The Pauli matrices transformed in the helicity basis can be written with the notation (\ref{polar}) as linear combinations of Pauli matrices
\be
\bar \sigma_x&=& {1\over |\Sigma|}
\begin{pmatrix}
\Sigma \cos\varphi
&
-i |\Sigma|\sin\varphi-{\Sigma_z }\cos\varphi
\cr 
i |\Sigma|\sin\varphi-{\Sigma_z }\cos\varphi
&
-\Sigma \cos\varphi
\end{pmatrix}
\nonumber\\
&=&-{\Sigma_z\over |\Sigma|}\cos\varphi\sigma_x+\sin\varphi\sigma_y
+{\Sigma\over |\Sigma|}\cos\varphi \sigma_z
\nonumber\\
\bar \sigma_y&=& {i\over |\Sigma|}
\begin{pmatrix}
-i\Sigma \sin\varphi
&
i \Sigma_z\sin\varphi+|\Sigma|\cos\varphi
\cr 
i \Sigma_z\sin\varphi-|\Sigma|\cos\varphi
&
i\Sigma \sin\varphi
\end{pmatrix}
\nonumber\\
&=&-{\Sigma_z\over |\Sigma|}\sin\varphi\sigma_x-\cos\varphi\sigma_y
+{\Sigma\over |\Sigma|}\sin\varphi \sigma_z
\nonumber\\
\bar \sigma_z&=& {1\over |\Sigma|}\begin{pmatrix}{\Sigma_z}&{\Sigma }\cr
  {\Sigma} &-\Sigma_z\end{pmatrix}
={\Sigma\over |\Sigma|} \sigma_x+{\Sigma_z \over |\Sigma|} \sigma_z.
\ee
This allows us to transform the velocity operator in the helicity basis
\be
\bar {\hat {v_i}}=\partial_{i} \epsilon+\begin{pmatrix}\left ( {\Sigma^2\over |\Sigma|}\partial_{i}{\Sigma_z\over \Sigma}\right )\cr - \Sigma\partial_{i}\varphi\cr \partial_{i}|\Sigma|\end{pmatrix}\cdot \V \sigma
\ee 
where $\partial_{i}$ denotes the derivative with respect to the i-th component
of the momentum. The current would be
\be
J_i=\frac e 2{\rm Tr} [\bar {\hat {v_i}} \delta \bar \rho]
\label{currd}
\ee
where we need the linearized solution of (\ref{hel1}) in the helicity basis with respect to an external electric field. Employing $(\partial U^+) U=-U^+\partial U$ one transforms
\be
U^+(\partial f) U=\partial \bar f+[U^+\partial U,\bar f]_-
\ee
such that the kinetic equation  (\ref{hel1}) reads
\be
\partial_t \bar \rho+[U^+\partial_t U,\bar \rho]_-+e \V E\cdot \V\partial_p \bar \rho&+&e
\V E\cdot [U^+\V \partial_p U,\bar \rho]_-\nonumber\\&
+&i[\bar H,\bar \rho]_-=0.
\ee
The corresponding linearized kinetic equation $ {\bar \rho}=\bar
\rho_0+\delta {\bar \rho}$ takes the same form 
\ba
&\partial_t \delta \bar \rho+[U^+\partial_t \delta U,\bar \rho_0]_-+e \V E\cdot \V \partial_p \bar \rho_0+e
\V E\cdot [U^+\V \partial_p U,\bar \rho_0]_-\nonumber\\&
+i[\bar H,\delta \bar \rho]_-=0.
\label{link}
\end{align}
For further use we neglect the time-dependence $\partial_t \delta U$ due to
selfconsistent mean field in the basis. The selfconsistently induced meanfield
term $[\delta {\bar H},\bar \rho_0]$ can be considered more convenient in the
next paragraph in spinor representation 
by the solution in the spin basis in the next part of this series.

With the help of the diagonal Hamiltonian (\ref{hbar}) one has for any matrix $A=\{a_{ij}\}$
\be
\left [\begin{pmatrix}\epsilon_+&0\cr 0 &\epsilon_-\end{pmatrix}, A\right ]&=&(\epsilon_+-\epsilon_-)\begin{pmatrix}0&a_{12}\cr -a_{21} &0\end{pmatrix}
\nonumber\\&=&\frac 1 2 (\epsilon_+-\epsilon_-)[\sigma_z,A]
\ee 
such that we can write (\ref{link}) with the equilibrium
distribution (\ref{F0})
\ba
&\partial_t \delta \bar \rho+e \V E\cdot \V \partial_p \bar \rho_0+
\frac 1 2 (f_+-f_-)e\V E \cdot [U^+\V \partial_p U,\sigma_z]_-\nonumber\\&
-{i\over 2}(\epsilon_+-\epsilon_-)[\delta \bar \rho ,\sigma_z]_-=0.
\end{align}
Consequently, the first part describes the diagonal response and the second part the response due to 
off-diagonal or band interference. The solution reads explicitly
\ba
\delta \bar \rho&=&-i {e \V E}\cdot 
\begin{pmatrix}
  {\V \partial_p f_+\over \omega}
&
{2 g\over \Delta \epsilon-\omega} \bra +\V \partial_p \ket -
\cr
{2 g\over \Delta \epsilon+\omega} \bra -\V \partial_p \ket +
 &
  {\V \partial_p f_-\over \omega}
\end{pmatrix}
\label{offdiag}
\end{align}
with $2 g=f_+-f_-$ and $\Delta \epsilon=\epsilon_+-\epsilon_-=2 |\Sigma|$.
The diagonal part leads to the standard dynamical Drude conductivity if a
scattering with impurities is considered $\omega\to  \omega+i/\tau$. The
second part is the reason for the anomalous Hall effect which we consider in the following.
%

With the help of 
(\ref{pm}) one has explicitly
\be
\bra \pm \partial\ket \mp&=&-i{\Sigma\over 2 |\Sigma|} \partial
\varphi\mp{\Sigma^2\over 2 |\Sigma|^2} \partial \left ({\Sigma_z\over \Sigma}\right )
\nonumber\\
\bra \pm \partial\ket \pm&=&-i{|\Sigma|\pm \Sigma_z\over 2 |\Sigma|} \partial \varphi
\label{derivh}
\ee
and the off-diagonal parts of (\ref{offdiag}) can be expressed as
\ba
\delta \bar \rho^{\rm AH}=
{\Sigma\over |\Sigma|}{g e \V E \cdot \over \omega^2\!-\!(\Delta \epsilon)^2}
&\biggl [
\left .
  \Delta \epsilon \left ( \V \partial_p \varphi \sigma_x \!+\! {\Sigma\over |\Sigma|} \V \partial_p\left ({\Sigma_z\over \Sigma}\right )\sigma_y\right )\right .
\nonumber\\
&\left .
+i\omega          \left ( \V \partial_p\varphi \sigma_y \!-\! {\Sigma\over |\Sigma|} \V \partial_p\left ({\Sigma_z\over \Sigma}\right )\sigma_x\right )
\right ].
\end{align}
The current (\ref{currd}) reads then
\be
J_\alpha=\sigma_{\alpha\beta} E_\beta
\ee
with the two parts of conductivity
\ba
&\sigma_{\alpha\beta}
={e^2\over 2}\sum\limits_p {\Sigma^2\over |\Sigma|}{2 g \over \omega^2-4|\Sigma|^2}
\nonumber\\
&\times
\left [
  2 \Sigma \left ( 
\partial_\beta\varphi \partial_\alpha\left ({\Sigma_z\over \Sigma}\right )-
\partial_\alpha\varphi \partial_\beta\left ({\Sigma_z\over \Sigma}\right )
\right )\right .
\nonumber\\
&\left .
-i\omega \left (
\partial_\beta\varphi \partial_\alpha\varphi+
{\Sigma^2 \over |\Sigma|^2}\partial_\beta\left ({\Sigma_z\over \Sigma}\right )\partial_\alpha\left ({\Sigma_z\over \Sigma}\right ) 
\right )
\right ].
\label{Hall}
\end{align}

\subsubsection{Dynamical asymmetric part}

The first part of (\ref{Hall}) is the standard anomalous Hall effect since it represents an asymmetric matrix noting 
\be
\partial_\alpha a\partial_\beta b-\partial_\beta a\partial_\alpha b=\epsilon_{\alpha\beta\gamma} (\V \partial a\times \V \partial b)_\gamma.
\ee
To simplify this asymmetric part further we perform the derivatives explicitly
\be
\V\partial \left ({\Sigma_z\over \Sigma}\right )\times \V \partial \varphi=-{1\over \Sigma^3}\epsilon_{ijk}\Sigma_i \V \partial\Sigma_j\times\V \partial \Sigma_k
\ee
and the first asymmetric part of (\ref{Hall}) can be written
\be
\sigma_{\alpha\beta}^{\rm as}={e^2\over 2}\sum\limits_p {g  \over
  1-{\omega^ 2\over 4|\Sigma|^2}}\,
\V e\cdot (\partial_\alpha \V e\times \partial_\beta \V e)
\label{asHall}
\ee
with $\V e =\V \Sigma/|\Sigma|$.
This describes the dynamical anomalous Hall conductivity as we can verify by the comparison with the dc Hall conductivity from the Kubo formula in the next section.

\subsubsection{Dynamical symmetric part}

The second symmetric part of (\ref{Hall}) is a pure dynamical conductivity and can be rewritten in a compact form
as well by noting
\be
&&\partial_\beta\varphi \partial_\alpha\varphi+
{\Sigma^2 \over |\Sigma|^2}\partial_\beta\left ({\Sigma_z\over \Sigma}\right )\partial_\alpha\left ({\Sigma_z\over \Sigma}\right )
\nonumber\\&&=\left (\partial_\alpha \V \Sigma \cdot \partial_\beta \V \Sigma-\partial_\alpha
  |\Sigma|\partial_\beta |\Sigma| \right ){1\over \Sigma^2}
\nonumber\\&&=\left (\partial_\alpha \V e \cdot \partial_\beta \V e \right
){|\Sigma|^2\over \Sigma^2}
\ee
with $\V e=\V \Sigma/|\V \Sigma|$.
Therefore we obtain
\be
\sigma_{\alpha\beta}^{\rm sym}=i{e^2\over 2} \sum\limits_p {{\omega\over
    2|\Sigma|}
g  \over
  1-\frac{\omega^2}{4|\Sigma|^2}} 
\,
\partial_\alpha \V e\cdot  \partial_\beta \V e.
\label{symHall}
\ee

\subsection{Anomalous Hall conductivity from Kubo formula}

For the reason of comparison we re-derive this results from the Kubo formula and consider the dc limit of the interband conductivity with band energies $\epsilon_n(p)$ and occupations $f_n(p)$. Due to band polarizations one has a finite current. The Kubo-Bastin-Streda formula reads ($\sum f_n=n_n$)
\be
\sigma_{\alpha\beta}={e^2\hbar \over i} \sum\limits_{nm}\sum\limits_p {f_m-f_n\over (\epsilon_n-\epsilon_m)^2} v_{nm}^\alpha v_{mn}^\beta
\label{kubo}
\ee
with the velocity in band basis
\be
\V v_{nm}&=&\bra n \hat{\V v}\ket m={1\over i\hbar}\bra n[\hat {\V x},\hat H]\ket m=\bra n [\V\partial_p, \hat H]\ket m
\nonumber\\
&=&\bra n \V\partial_p\ket m (\epsilon_m-\epsilon_n).
\ee
The conductivity becomes therefore with the notation $(\V \partial_p)_\alpha=\partial_\alpha$
\be
\sigma_{\alpha\beta}
&=&-{e^2\hbar \over i} \sum\limits_{nm}\sum\limits_p (f_m-f_n)\bra n \partial_\alpha\ket m\bra m \partial_\beta\ket n
\nonumber\\
&=&{e^2\hbar \over i} \sum\limits_{n}\sum\limits_p f_n\bra n  \partial_\alpha\partial_\beta-\partial_\beta\partial_\alpha\ket n
\nonumber\\
&=&\epsilon_{\alpha\beta \gamma}{e^2\hbar \over i} \sum\limits_{n}\sum\limits_p f_n\bra n  (\V\partial_p\times \V\partial_p)_\gamma\ket n
\nonumber\\
&=&\epsilon_{\alpha\beta \gamma}{e^2\hbar \over i} \sum\limits_{n}\sum\limits_p f_n  (\V\partial_p \times \bra n\V\partial_p\ket n )_\gamma
\nonumber\\
&=&-\epsilon_{\alpha\beta \gamma}{e^2} \sum\limits_{n}\sum\limits_p f_n  (\V\partial_p \times \V a_n)_\gamma
\ee
where we introduced in the last step the Berry-phase connection
\be
\V a_n=i \hbar \bra n\V\partial_p\ket n=\bra n\V x\ket n  
\ee
and $\V\partial \times \V a_n$ is the Berry-phase curvature.

Now we specify this formula for the two-spin band problem where the Berry phase connection with the help of (\ref{derivh}) reads
\be
\V a_\pm=i \hbar \bra \pm \V\partial_p\ket \pm=\hbar {\Sigma\pm\Sigma_z\over 2\Sigma} \V\partial_p \varphi
\ee
and the Berry curvature
\be
\V \partial_p\times\V a_\pm=\mp {\hbar \over 2 \Sigma^3} \epsilon_{ijk}\Sigma_i \V \partial_p \Sigma_j\times\V\partial_p \Sigma_k
\ee
or
\be
(\V \partial_p\times\V a_\pm)_\gamma&=&\mp {\hbar \over 2 \Sigma^3} \epsilon_{\alpha\beta\gamma}\epsilon_{ijk}\Sigma_i \partial_\alpha \Sigma_j\partial_\beta \Sigma_k
\nonumber\\
&=&\mp {\hbar \over 2 \Sigma^3} \epsilon_{\alpha\beta\gamma}\V \Sigma \cdot (\partial_\alpha \V \Sigma \times \partial_\beta \V \Sigma).
\ee
Therefore the dc Hall conductivity reads finally
\be
\sigma_{\alpha\beta}^{\rm dc}={e^2\hbar\over 2}\sum\limits_p g\,\,
\V e\cdot (\partial_\alpha \V e\times \partial_\beta \V e)
\ee
which is exactly the dc limit of (\ref{asHall}).

\subsection{Spin-Hall and anomalous Hall effect in spin basis}

\subsubsection{Anomalous and inverse Hall effect}
Now we solve the equations (\ref{homog}) once more directly in the spin basis. This has the advantage that the ambiguous term $U^+\partial_tU$ in the helicity basis does not appear and in this way we will see that it does not contribute to the final result. Moreover we have the relaxation time in the kinetic equation which means that in the end we can understand $\omega\to\omega+i/\tau$. In order to keep the comparison as near as possible to the above two ways of derivation we  keep $\omega$ and shift in the end. 

Equations (\ref{homog}) are decoupled since we neglect mean-field effects and magnetic fields. Linearizing and noting that $\V \Sigma\times \V g=0$ since $\V g=\V e (f_+-f_-)/2$ we obtain after the Fourier transform of time
\be
\delta f(\omega,p)&=&-{i\over \omega} eE\p p f\nonumber\\
\delta \V g(\omega,p)&=& 
{i \omega \over 4 |\Sigma|^2-\omega^2} e E\partial_p\V g
\nonumber\\&&
-4 i{1\over \omega (4 |\Sigma|^2-\omega^2)}\V \Sigma (\V \Sigma \cdot eE\partial_p \V g)
\nonumber\\&&
-2  {1  \over 4 |\Sigma|^2-\omega^2} \V \Sigma \times eE\partial_p \V g.
\label{deltarho}
\ee 
With the help of
\be
{1\over 4 |\Sigma|^2-\omega^2}\begin{pmatrix} -i \omega \cr 
{4 i|\Sigma|^2\over \omega} \cr 2 |\Sigma| \end{pmatrix}
=\int\limits_0^\infty {\rm e}^{i\omega t} \begin{pmatrix} \cos 2|\Sigma| t\cr 1-\cos 2|\Sigma| t \cr \sin 2|\Sigma t|\end{pmatrix},
\label{costime}
\ee
one sees that each of the terms in (\ref{deltarho}) correspond to a specific precession motion analogously to the one seen in the conductivity of a charge in crossed electric and magnetic fields 
\ba
&\V J(t)=\sigma_0 \int\limits_0^t {d\bar t\over \tau} {\rm e}^{-{\bar t\over \tau}}\left \{ \cos{(\omega_c \bar t)}\V E(t-\bar t)
\right .\nonumber\\&\left .
\!+\! \sin{(\omega_c \bar t)}\V E(t\!-\!\bar t)\times \V B_0
\!+\![1\!-\!\cos{(\omega_c \bar t)}][\V E(t\!-\!\bar t)\cdot \V B_0]\V B_0\right \}
\label{jtime}
\end{align}
as the solution of the Newton equation of motion
\be
m_e \dot {\V v}=e (\V v\times \V B)+e \V E-m_e {\V v\over \tau}.
\label{start}
\ee
It illustrates the threefold orbiting of the electrons with cyclotron frequency: (i) in the direction of the electric and (ii) magnetic field, and (iii) in the direction perpendicular to the magnetic and electric field.

The charge current (\ref{current}) or (\ref{currents})
consists of the normal current as the first part,
\be
e \sum\limits_p \p \alpha \epsilon \delta f={i n e^2\over \omega^+} E_\alpha= {i n e^2 \tau\over 1-i\omega \tau} E_\alpha 
\ee
 and the anomalous current due to spin-polarization as the second part of (\ref{currents}). The latter one
contains the standard anomalous Hall effect as the third term of (\ref{deltarho}) and their first and second term will combine together to the symmetric part of the anomalous conductivity as we will demonstrate now. Since $\V \Sigma \cdot \partial \V e=0$, the third term of (\ref{deltarho}) leads to
\be
\sigma_{\alpha\beta}^{\rm as}&=&-e^2 \sum\limits_p \partial_\alpha \V \Sigma(\V \Sigma\times \partial_\beta\V e){f_+-f_-\over 4 |\Sigma|^2-\omega^2}
\nonumber\\
&=&-{e^2\over 2} \sum\limits_kp {g\over 1-{\omega^ 2\over 4|\Sigma|^2}}\partial_\alpha \V e(\V e\times \partial_\beta\V e)
\ee
and one recognizes the anomalous Hall conductivity (\ref{asHall}).

The first and second term of (\ref{deltarho}) combine together with the symmetric part
\ba
\sigma_{\alpha\beta}^{\rm sym}&={i e^2\over 2 \omega } \sum\limits_p {2 \partial_\alpha \V \Sigma\over 4 |\Sigma|^2\!-\!\omega^2}
\biggl [
\omega^2 (g \partial_\beta \V e\!+\!\V e \partial_\beta g)
\!-\!4 \V \Sigma |\Sigma|\partial_\beta g 
\biggr ]
\nonumber\\
&={i e^2 \over 2 \omega } \sum\limits_p {2\over 4 |\Sigma|^2\!-\!\omega^2}
\biggl [{\omega^2 g\over |\Sigma|} ( \partial_\alpha \V \Sigma\partial_\beta \V \Sigma\!-\!\partial_\alpha |\Sigma|\partial_\beta |\Sigma|)
\nonumber\\&
\qquad\qquad\qquad\qquad +\omega^2\partial_\alpha|\Sigma| \partial_\beta g-4 |\Sigma|^2 \partial_\alpha|\Sigma|\partial_\beta g\biggr ]
\nonumber\\
&={i \omega e^2\over 4} \sum\limits_p {g\over 1-{\omega^2\over 4 |\Sigma|^2}}
{1\over |\Sigma|^3}\biggl ( \partial_\alpha \V \Sigma\partial_\beta \V \Sigma\!-\!\partial_\alpha |\Sigma|\partial_\beta |\Sigma|\biggr )
\nonumber\\&
\quad + {i e^2 \over \omega} \sum\limits_p g \partial_\alpha\partial_\beta |\Sigma|
\label{as1}
\end{align}
where we have used in the second step the relation
\ba
\partial_\alpha \V \Sigma\partial_\beta \V {\Sigma\over |\Sigma|}={1\over |\Sigma|}\biggl ( \partial_\alpha \V \Sigma\partial_\beta \V \Sigma\!-\!\partial_\alpha |\Sigma|\partial_\beta |\Sigma|\biggr )=|\Sigma|\partial_\alpha \V e \cdot \partial_\beta \V e.
\end{align}
Since the last term in (\ref{as1})  vanishes due to symmetry in $p$ we obtain exactly (\ref{symHall}).

Summarizing, the total charge current (\ref{currents}) is given by 
\be
J_\alpha=\sigma^{\rm D} E_\alpha+(\sigma_{\alpha\beta}^{\rm as}+\sigma_{\alpha\beta}^{\rm sym})E_\beta
\label{currt1}
\ee
with the usual Drude conductivity $\sigma^{\rm D}=ne^2 \tau/m_e$ and
the symmetric and asymmetric parts of the anomalous Hall conductivity (\ref{asHall}) and (\ref{symHall})
\be
\left .
\begin{matrix}
\sigma_{\alpha\beta}^{\rm as}
\cr\cr
\sigma_{\alpha\beta}^{\rm sym}
\end{matrix} 
\right \}={e^2\over 2}\sum\limits_p {g  \over
  1-{\omega^ 2\over 4|\Sigma|^2}}\,\left \{ \begin{matrix}
\V e\cdot (\partial_\alpha \V e\times \partial_\beta \V e)
\cr\cr
{i\omega\over
    2|\Sigma|}\partial_\alpha \V e\cdot  \partial_\beta \V e
\end{matrix} \right .
\label{145}
\ee
and $\V e =\V \Sigma/|\Sigma|$. Note that from our kinetic equation with the relaxation time approximation we understand the above formulas as $\omega\to \omega+i/\tau$ which leads in the static limit the modifications of the Kubo expression due to collisions.

For zero temperature and linear Rashba spin-orbit coupling we can integrate these expressions analytically. We consider the electric field in x-direction and obtain
\ba
&\sigma^{\rm as}_{yx}={e^2\over 4 \pi \hbar} \Sigma_n \tau_\omega \arctan{\left [
{2 \epsilon_\beta \tau_\omega\over \hbar^2+4 (2 \epsilon_\beta \epsilon_F+\Sigma_n^2)\tau_\omega^2}
\right ]}
\nonumber\\
&\to{e^2\over 4 \pi \hbar}\left \{
\begin{matrix}
{\epsilon_\beta \Sigma_n\over 2 \epsilon_\beta \epsilon_f+\Sigma_n^2} & \omega=0,\tau\to\infty
\cr
{\Sigma_n\over \omega} {\rm artanh}\left [{2 \epsilon_\beta \omega \over \hbar^2\omega^2-4 (2 \epsilon_\beta \epsilon_F+\Sigma_n^2)}\right ]
& \omega\ne 0,\tau\to\infty
\end{matrix}
\right .
\label{146}
\end{align}
with the Rashba energy $\epsilon_\beta=m\beta_R^2/\hbar$ and the dynamical result is given $\tau_\omega= \tau/(1-i \omega \tau)$ in (\ref{146}). We see that the anomalous Hall effect vanishes with vanishing effective Zeeman field
\be
\Sigma_n=|n \V V+\V s V_0+\mu_B \V B|.
\label{147}
\ee

\begin{figure}
\includegraphics[width=8cm]{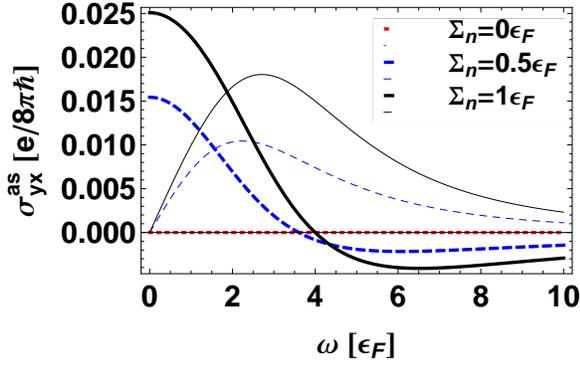}
\caption{\label{anomHall_xyz(wb)}
The dynamical anomalous Hall conductivity (\ref{146}) vs. frequency for different values of the effective Zeeman field (\ref{147}) and a relaxation time $\tau=0.3\hbar\epsilon_F$ and a Rashba energy of $m \beta_R^2=0.1 \epsilon_F$. Real parts are thick and imaginary are thin lines.}
\end{figure}

We see from figure \ref{anomHall_xyz(wb)} that the anomalous Hall conductivity is strongly dependent on the frequency showing even a sign change at high frequencies which had been reported before \cite{Zhou:2006}. This sign change is connected with a collisional damping which means that it is suppressed by collisions. This damping as expressed by the imaginary part of the conductivity vanishes in the static limit.

\begin{figure}
\includegraphics[width=8cm]{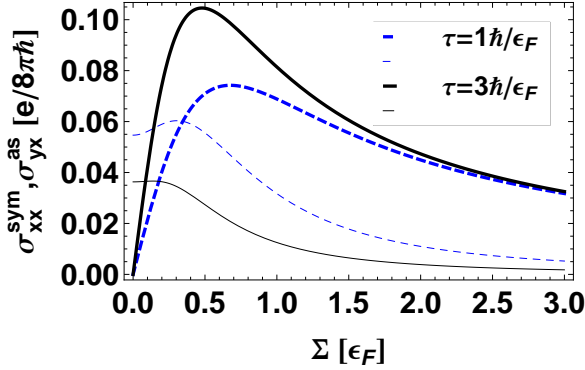}
\caption{\label{anomHall_xxyz(0s)}
The static anomalous Hall conductivity of (\ref{146}) (thick) and inverse Hall conductivity (\ref{148}) (thin) vs. effective Zeeman field for two different relaxation times and a Rashba energy of $m \beta_R^2=0.1 \epsilon_F$.}
\end{figure}

The absolute value is dependent on the effective Zeeman field (\ref{147}) as one sees from the scaling of the static limit plotted in figure \ref{anomHall_xxyz(0s)}. The static anomalous Hall conductivity possesses a maximum at certain Zeeman terms which are dependent on the relaxation time.

The symmetric part of (\ref{145}) in fact yields $\sigma_{yx}^{\rm sym}=0$ and
\ba
\sigma^{\rm sym}_{xx}&={e^2\over 16 \pi \hbar} \left \{
{4 \epsilon_\beta \Sigma_n^2 \tau_\omega\over 2 \epsilon_\beta \epsilon_f+\Sigma_n^2}
\right .\nonumber\\ &+\left .
(1-4 \Sigma_n^2 \tau_\omega^2)
\arctan{\left [
{4 \epsilon_\beta \tau_\omega\over \hbar^2+4 (2 \epsilon_\beta \epsilon_F+\Sigma_n^2)\tau_\omega^2}
\right ]}
\right \}
\nonumber\\
&\to{e^2\over 4 \pi \hbar}\left \{
\begin{matrix}
0 & \omega=0,\tau\to\infty
\cr
{\rm imaginary}
&\omega\ne 0,\tau\to\infty
\end{matrix}
\right .
\label{148}
\end{align}
which shows that it represents a contribution in the direction of the applied electric field and is caused by collisional correlations. We interpret it as an inverse Hall effect. This dynamical result is different from the spin accumulation found in \cite{BK06} basically by the $arctan$ term and therefore no sharp resonance feature. Expanding, however, in small spin-orbit coupling
\ba
\sigma^{\rm sym}_{xx}&={e^2\over 2 \pi \hbar} \,{\epsilon_\beta \tau\over1+4 \Sigma_n^2 \tau^2}+o(\epsilon_\beta^2)
\end{align}
shows that the static limit agrees with \cite{BK06,IBM03}. Please note that if one sets $\Sigma_n\to 0$ before expanding a factor 1/2 appears which illustrates the symmetry breaking by the effective Zeeman term.

In figure \ref{anomHall_xxyz(0s)} we compare this expression with the anomalous Hall conductivity for two different relaxation times. While the static anomalous Hall conductivity vanishes with the effective Zeeman field the inverse Hall effect remains finite which value is easily seen from (\ref{148}). Both the anomalous Hall conductivity as well as the inverse Hall conductivity possess a maximum at certain effective Zeeman fields.

\begin{figure}
\includegraphics[width=8cm]{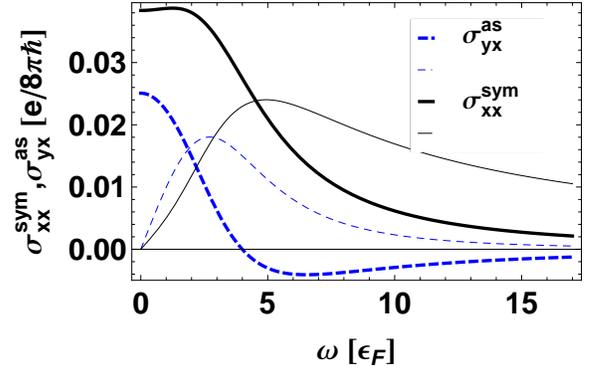}
\caption{\label{anomHall_xxyz(ws)}
The comparison of the dynamical anomalous Hall conductivity (\ref{146}) (solid) and the inverse Hall conductivity (\ref{148}) (dashed) for a relaxation time $\tau=0.3\hbar\epsilon_F$, a Rashba energy of $m \beta_R^2=0.1 \epsilon_F$ and an effective Zeeman energy $\Sigma_n=1\epsilon_F$. The real parts are thick lines, and the imaginary parts are thin lines.}
\end{figure}

The comparison of the dynamical anomalous Hall effect and the inverse Hall effect finally can be found in figure \ref{anomHall_xxyz(ws)}. In contrast to the anomalous Hall effect the inverse Hall effect does not show a sign change which means the current remains in the direction of the applied electric field as it should.

\subsubsection{Spin-Hall effect}

Now we can consider the spin current (\ref{currents}) with the help of the long-wavelength solution (\ref{deltarho}) without magnetic fields. The part with $\delta f$ vanishes after partial integration and symmetry in $p$. The other terms group  into a normal spin-current and an (anomalous) part representing the spin-Hall effect
\be
\V S_\alpha=-{e \tau \over m_e(1-i\omega \tau)}\V s E_\alpha+\V \sigma_{\alpha\beta} E_\beta.
\ee
The spin-Hall coefficient consists analogously as the anomalous Hall effect of a symmetric and an asymmetric part ($\omega\to \omega+i/\tau$)
\be
\left .
\begin{matrix}
\V \sigma_{\alpha\beta}^{\rm as}
\cr\cr
\V \sigma_{\alpha\beta }^{\rm sym}
\end{matrix} 
\right \}={e\over m_e\omega} \sum\limits_p {p_\alpha g  \over
  1-{\omega^ 2\over 4|\Sigma|^2}}\,\left \{ \begin{matrix}
{i\omega\over
    2|\Sigma|} \V e\times \partial_\beta \V e
\cr\cr
i\partial_\beta \V e
\end{matrix} \right ..
\label{spinHall}
\ee

We see that both effects the spin-Hall effect and the anomalous Hall effect appear without magnetic fields and have their origin in the anomalous parts of the currents due to spin-orbit coupling. Though the normal and anomalous currents exactly compensate in the stationary state, the current due to a disturbing external electric field shows a finite asymmetric and symmetric part (\ref{spinHall}) not known in the literature.

Explicit integration of (\ref{spinHall}) is possible to carry out in zero temperature and linear Rashba and Dresselhaus coupling. We assume the electric field in the x-axis and we get for the Rashba linear spin-orbit coupling
\ba
\sigma_{yx}^z\!=\!{e\over 8 \pi \hbar}\!\left [
1\!-\!{1\!+\!4 \Sigma_n^2 \tau_\omega^2\over 4 \epsilon_\beta \tau_\omega} 
\arctan{\!\left (\!
4\hbar \epsilon_\beta \tau_\omega\over \hbar^2\!+\!4 \tau_\omega^2 (2 \epsilon_\beta \epsilon_F\!+\!\Sigma_n^2)
\!\right )}
\!\right ]
\label{SHR}
\end{align}
with $\tau_\omega= \tau/(1-i \omega \tau)$.
Only the z-axis survives in linear spin-orbit coupling.
Neglecting the selfenergy and using the static limit it is just the result of \cite{Schliemann:2004,Sch06}. 

The so-called universal limit appears if one takes the limit of vanishing collision frequency
\be
\sigma_{yx}^z={e\over 8 \pi \hbar} \, {2\epsilon_\beta \epsilon_f\over 2\epsilon_\beta \epsilon_f+\Sigma_n^2}+o(1/\tau).
\ee 
We see how the selfenergy including the Zeeman term (\ref{147}) modifies this
"universal limit" which already hints at questioning of this notion.

For the Dresselhaus linear spin-orbit coupling we obtain just (\ref{SHR}) with opposite sign. Therefore if we approximate the combined effect by adding the two specific results we see that the constant term vanishes and the difference of the expressions with corresponding Rashba and Dresselhaus energies occur. The correct treatment of both couplings together leads to involved angular integrations and escape analytical work.

The universal constant $e/8\pi\hbar$ has been first described by
\cite{SCNSJD04} and raised an intensive discussion. It was shown that the
vertex corrections cancel this constant \cite{IBGML04,RS05}. A suppression of
Rashba spin-orbit coupling has been obtained due to disorder \cite{CL05}, or
electron-electron interaction \cite{D05} and found to disappear in the
self-consistent Born approximation \cite{Liu06}. The conclusion was that the
two dimensional Rashba spin-orbit coupling does not lead to a spin-Hall effect
as soon as there are relaxation mechanisms present which damp the spins
towards a constant value. In order to include such effects one has to go
beyond meanfield and relaxation-time approximation by including vertex
corrections \cite{ZNA01}. The spin-Hall effect does not vanish with magnetic fields or spin-dependent scattering processes \cite{Sch06}. 

Please note that the universal constant in (\ref{SHR}) is necessary to obtain the correct small spin-orbit coupling result
\be
\sigma_{yx}^z={e\over \pi \hbar} {\epsilon_f \tau^2\over (1-i \omega \tau)^2+4 \Sigma_n^2 \tau^2} \epsilon_\beta+o(\epsilon_\beta^2).
\ee
Without the Zeeman term $\Sigma_n\to 0$ and for small spin-orbit coupling this
agrees with the dynamical result of \cite{BK06} where the definition of spin
current has been employed in terms of physical argumentation. Again the result
here differs from the resonant structure found in \cite{BK06} by the $arctan$ term but the static limit agrees with the result of \cite{Schliemann:2004,Sch06}. 

The dynamical result (\ref{SHR}) describes the influence of an external
magnetic field as well as mean magnetizations due to magnetic impurities. The
advantage of the result here is the simplicity in which the frequency dependence enters and the combined effect of external magnetic field, spin polarizations and mean magnetization described by one vector selfenergy (\ref{sig}) called the effective Zeeman field. 

The z-component of the spin-Hall coupling in the x-direction becomes a modified coefficient (\ref{SHR})
\be
\sigma_{xx}^z={2\over \hbar} \Sigma_n \tau\sigma_{xy}^z.
\label{SHR1}
\ee
The medium effects or magnetic field or magnetization condensed in (\ref{sig}) triggers a second spin-Hall direction in plane with the z-axes and the electric field. This observation is interpreted as the inverse spin-Hall effect. The here presented inverse spin-Hall effects are the underlying physics in the recently observed terahertz spin signals \cite{KBMENMZFMBWROM13} in magnetic heterostructures.

\begin{figure}
\includegraphics[width=8cm]{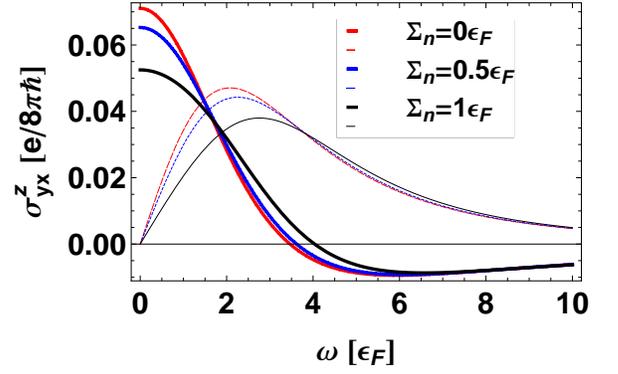}
\caption{\label{spinHall_xyz(wb)}
The dynamical spin-Hall coefficient vs frequency for different values of the effective Zeeman field and a relaxation time $\tau=0.3\hbar\epsilon_F$ and a Rashba energy of $m \beta_R^2=0.1 \epsilon_F$, real parts are shown as thick and imaginary parts as thin lines.}
\end{figure}

\begin{figure}
\includegraphics[width=8cm]{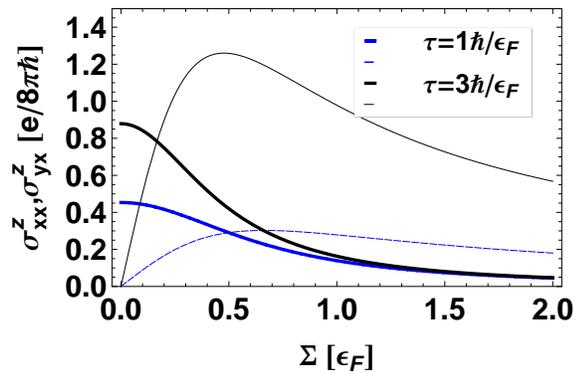}
\caption{\label{spinHall_xxyz(0s)}
The static spin-Hall coefficients of (\ref{SHR}) (thick) and (\ref{SHR1}) (thin) effective Zeeman field for two different relaxation times and a Rashba energy of $m \beta_R^2=0.1 \epsilon_F$.}
\end{figure}

In figure \ref{spinHall_xyz(wb)} we plot the dynamical spin-Hall coefficients for different effective Zeeman fields. 
The coefficients becomes suppressed with increasing Zeeman field as seen also in figure \ref{spinHall_xxyz(0s)} and with larger scattering frequency $1/\tau$. Interestingly the real part of the spin-Hall conductivity shows a sign change at a specific frequency. Though this effect is interesting this reversed current is strongly damped represented by the imaginary part.

The static limit becomes suppressed as seen in Fig. \ref{spinHall_xxyz(0s)} dependent on the relaxation time and the effective Zeeman field. The in-plane component (\ref{SHR1}) is zero for the absent Zeeman term and shows a characteristic maximum at certain effective Zeeman fields.

\begin{figure}
\includegraphics[width=8cm]{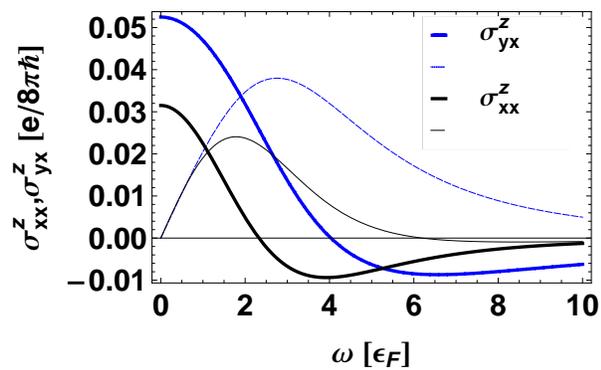}
\caption{\label{spinHall_xxyz(ws)}
The comparison of the dynamical spin-Hall coefficients (\ref{SHR}) (solid) and the inverse spin-Hall coefficient (\ref{SHR1}) (dashed) for a relaxation time $\tau=0.3\hbar\epsilon_F$, a Rashba energy of $m \beta_R^2=0.1 \epsilon_F$ and an effective Zeeman energy $\Sigma_n=1\epsilon_F$.}
\end{figure}

The dynamical spin Hall coefficient and the inverse spin Hall coefficient are compared in figure \ref{spinHall_xxyz(ws)}. We recognize a sign change at higher frequencies connected with a strong damping for both effects. Compared to the inverse Hall effect the inverse spin-Hall current can be directed both parallel and anti-parallel to the electric field.
 
It is important to note that the here presented linearized mean-field 
kinetic equation with a relaxation-time approximation is not capable to 
describe correctly the collisional correlations which has to be performed 
beyond the relaxation-time approximation. Especially the above-mentioned 
vertex corrections seem to lead to cancellations of the result. Here we 
restrict ourselves to a Drude level of conductivity.

\section{Summary}

The coupled kinetic equation for particle and spin polarization is derived including meanfields, spin-orbit coupling and arbitrary magnetic and electric field strength. This is achieved by using a gauge-invariant formulation and keeping the quantum spin structure  as commutators/anticommutators through gradient approximations. Both equations have the expected structure of a drift controlled by a mean quasiparticle velocity renormalized by meanfields and the effective Lorentz force which contains besides the magnetic field also the vector parts of the selfenergy. These equations are coupled exactly by the latter one. Additionally the polarization distribution exercises a precession around a direction given by this vector selfenergy. This latter one together with the magnetic field establishes an effective medium-dependent magnetic field and can be considered as a many-body extension of the Zeeman field. 
The stationary solution shows a unique splitting into two bands controlled by the vector selfenergy. 

Here the selfconsistent precession direction appears. For linear spin-orbit coupling and zero temperature an exact cancellation of spin-orbit coupling is found for the polarization.
 
We have calculated charge and spin currents and show that besides the regular currents an anomalous part appears due to spin-orbit coupling. The regular and anomalous currents compensate exactly in the stationary state. For transport with respect to an external electric field we calculate the spin-Hall coefficient as well as the anomalous Hall conductivity. Both consist of an asymmetric part which in the case of the anomalous Hall effect agrees with the standard one from the Kubo formula or Berry phases and an additional symmetric part interpreted as the inverse Hall effect with an expression not presented so far. The corresponding spin-Hall effects are described as well and are presented here in their dynamical form dependent on the magnetic field and the mean magnetization of the system.

A sign change of the Hall conductivity is reported for higher frequencies than a critical one given in terms of the relaxation time. The inverse Hall effect does not show such sign change in accordance with causality. The spin-Hall and inverse spin-Hall effects both show such sign change where the latter one is an expression of the non-conserved spin current.

In the second part of this paper series we will present the linear response results including the meanfield and magnetic field which allows us to discuss the dynamical response functions. From this we will see the collective modes and how the spin-orbit coupling influences the screening properties. 

\acknowledgments
Discussions with Janik Kailasvuori, Alvaro Ferraz and Ozgur Bozat
are gratefully mentioned.
Financial support by the Brazilian Ministry of Science 
and Technology is acknowledged.

\bibliography{bose,kmsr,kmsr1,kmsr2,kmsr3,kmsr4,kmsr5,kmsr6,kmsr7,delay2,spin,spin1,refer,delay3,gdr,chaos,sem3,sem1,sem2,short,cauchy,genn,paradox,deform,shuttling,blase,spinhall,spincurrent,tdgl,pattern,zitter,isospin}

\end{document}